\documentclass[a4paper,fleqn]{cas-dc}
\usepackage[T1]{fontenc}
\usepackage{palatino}
\usepackage{lipsum}
\usepackage{microtype}
\usepackage[numbers,sort&compress]{natbib}
\usepackage{siunitx}
\usepackage{amsmath}
\usepackage{amsfonts}
\usepackage{amssymb}
\usepackage{array}
\usepackage{graphicx}
\usepackage{tabularx}
\usepackage{multirow}
\usepackage{multicol}
\usepackage[version=4]{mhchem}
\usepackage{gensymb}
\usepackage{caption}
\usepackage{subcaption}
\usepackage{framed}
\usepackage{nomencl}
\makenomenclature

\usepackage{newfloat}

\usepackage{textcomp, gensymb}
\usepackage{comment}
\usepackage{amsmath,amssymb}
\usepackage[dvipsnames]{xcolor}
\usepackage{siunitx}
\usepackage{esvect}

\usepackage{lineno}
\DeclareSIUnit{\kwhe}{kWh_e}
\DeclareSIUnit{\kwh}{kWh}
\DeclareSIUnit{\crate}{C}

\DeclareFloatingEnvironment[
    name=Fig. A,
    fileext=lof
    placement={t}
]{figap}

\makeatletter
\renewcommand{\thefigap}{\arabic{figap}}
\renewcommand{\fnum@figap}{\textbf{Fig. A\thefigap}}
\makeatother

\graphicspath{{Fig_sup/}}

\begin{document}

\let\WriteBookmarks\relax
\def\floatpagepagefraction{1}
\def\textpagefraction{.001}
\shorttitle{Multifaceted thermal regulation in electrochemical batteries}
\shortauthors{Bozorg and Torres}

\title [mode = title]{Multifaceted thermal regulation in electrochemical batteries using cooling channels and foam-embedded phase change materials}

\author[1]{Mehdi V. Bozorg}
\credit{Conceptualisation, Investigation, Methodology, Roles/Writing - original draft}
\author[1]{Juan F. Torres}[orcid=0000-0002-3054-8638]
\credit{Funding acquisition, Conceptualisation, Supervision, Methodology, Writing - review \& editing}
\cormark[1]
\address[1]{ANU HEAT Lab, School of Engineering, The Australian National University, Canberra, Australia}

\cortext[cor1]{Corresponding author: felipe.torres@anu.edu.au}

\begin{abstract}
Lithium-ion batteries are widely used in electric vehicles and grid energy storage systems. Compared to cylindrical batteries, prismatic cells are the primary choice because of their advantage for dense packing. However, thermal runaway and temperature inhomogeneities are the main thermal regulation problems that affect their reliability, safety, and useful life. Here, we propose and assess a multifaceted cooling system composed of water channels (active cooling) and metallic foam embedded with two types of phase-change materials or PCMs (passive cooling) with different melting points. We show that a multifaceted thermal regulation strategy can improve both cooling effectiveness and temperature homogeneity through a representative prismatic battery module. Our numerical results indicate that for a battery pack cooled with a water channel (\SI{3}{\crate} discharge rate), a dual-PCM arrangement can reduce the maximum temperature by \SI{1.3}{\celsius} and \SI{2.7}{\celsius} compared to a mono-PCM arrangement and a battery pack without PCM, respectively. The maximum temperature difference within the cell is also \SI{1.2}{\celsius}. Therefore, multi-PCM thermal management systems show better performance than their mono-PCM predecessors in terms of lowering the maximum battery temperature and improving thermal homogeneity. This work motivates the development of multifaceted thermal management systems with active and passive cooling to improve the long-term performance of electrochemical battery cells.
\end{abstract}

\begin{keywords}
\sep lithium--ion battery \sep dual PCMs \sep metal foam \sep forced liquid cooling \sep computational fluid dynamics \sep thermal management system
\end{keywords}

\maketitle


\section{Introduction}

Global warming is the main threat to Earth and comes from excessive carbon emissions from fossil fuel consumption. Much research has targeted decarbonisation strategies to reduce greenhouse gas emissions by developing energy conversion technologies based on renewable resources that reduce our dependence on hydrocarbon fuels. Moreover, more than one-fifth of global carbon emissions are related to the transportation sector \cite{jelica2018hourly}. Therefore, significant efforts have been made by engineers, scientists, entrepreneurs, and governments to produce sustainable electric vehicles (EVs) \cite{goetzel2022empirical}. Today, hybrid electric vehicles (HEVs) and EVs are significantly promoting the development of private and public transportation systems to globalise green travel. The lithium ion (Li-ion) battery has become the industry standard for EVs/HEVs to power their electric motors \cite{bernagozzi2022heat}. In addition, the electricity generated by renewable resources could be stored in large-scale Li-ion batteries to feed electricity into the grid \cite{chen2022experimental}. For example, the surplus of electricity produced from off-grid photovoltaic systems during peak solar hours could be transferred to Li-ion battery packs and then used to charge EVs. The advantages of Li-ion batteries include a low self-discharge rate, relatively high energy density rates ($>$ \SI{180}{\watt\hour\per\kilo\gram}) \cite{lokhande2022bacterial}, large power-per-weight (\SI{1.5}{\kilo\watt\per\kilo\gram}) \cite{sui2022robust}, high efficiency, reliability, and a relatively long life cycle compared to other electrochemical batteries. However, large-scale lithium batteries still face numerous challenges, such as degradation \cite{liu2020janus}, safety issues due to instability at high temperatures (runaway) \cite{finegan2018identifying}, production with environmentally friendly materials \cite{liu2022dual}, and recyclability~\cite{jaguemont2016comprehensive}.

\begin{figure*}
    \centering
    \includegraphics[width=0.8\textwidth]{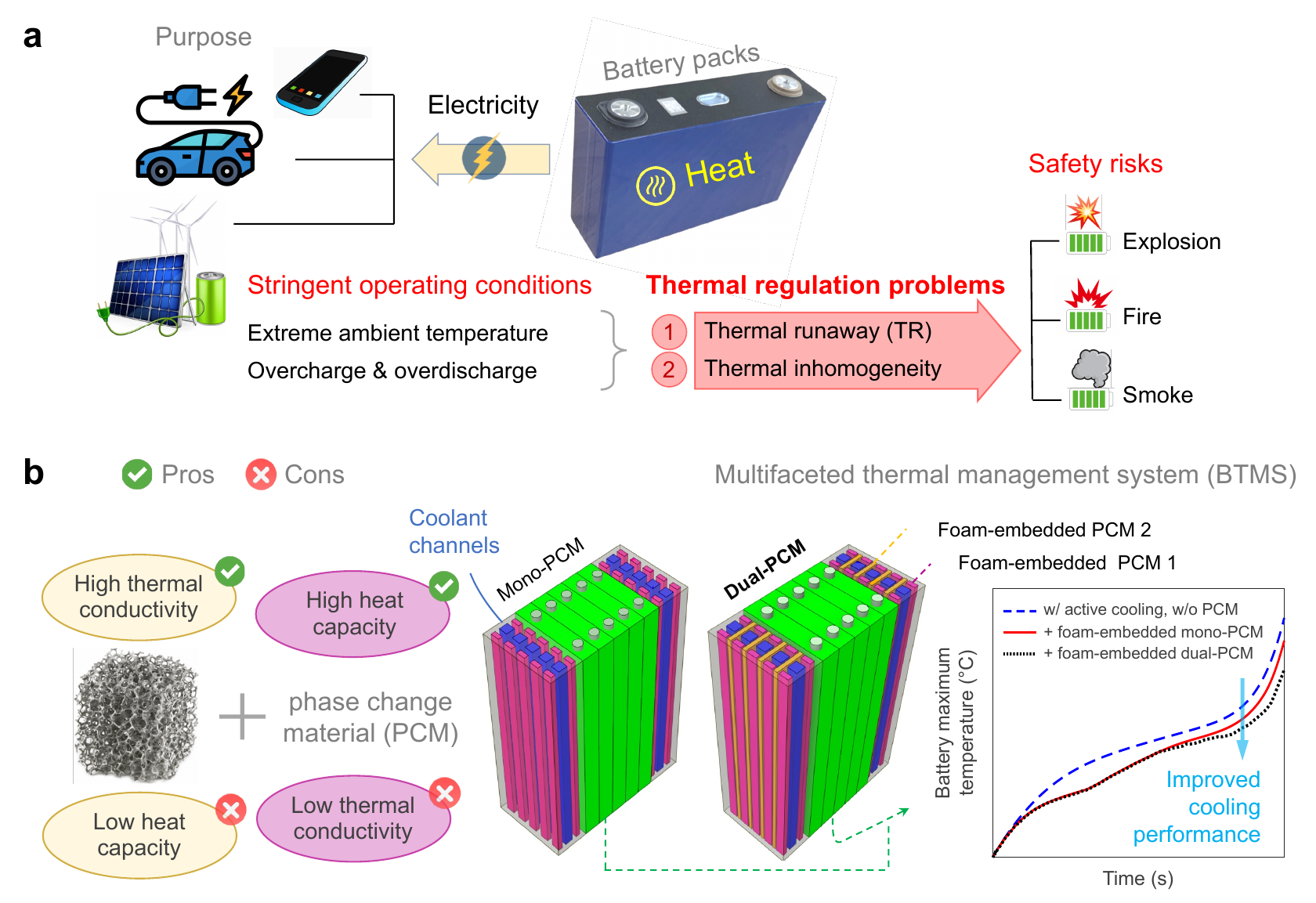}
    \caption{Concept of multifaceted thermal management strategies for electrochemical batteries.
    \textbf{(a)}~Applications of electrochemical batteries, their stringent conditions, thermal regulation problems including thermal runaway and inhomogeneity, as well as associated safety risks.
    \textbf{(b)}~Multifaceted cooling strategy for battery packs including advantages and disadvantages of each element. The proposed battery thermal management system is comprised of an active method using water channels, and a passive methods using dual foam-embedded PCMs. Our hypothesis is an improved cooling performance when implementing a multifaceted thermal management strategy.}
\label{fig:abstract}
\end{figure*}

The performance of Li-ion batteries is closely related to the maximum temperature of the module and the homogeneity of the temperature $(\Delta\textit{T} = \textit{T}_{\mathrm{max}} - \textit{T}_{\mathrm{min}})$ within the battery module. These temperatures are transient during the charge and discharge phases. Consequently, exceeding a specific temperature range deteriorates the reliability and safety of lithium ion batteries. Exothermic reactions that trigger fire and explosion \cite{mbulu2021experimental} and reduced lifetime due to cell degradation (loss of battery capacity during charge / discharge cycles) are the result of perturbations in electrochemical reactions whose magnitude is affected by the temperature of the cell \cite{pires2022assessment}. One of the most serious challenges of Li-ion batteries is thermal runaway (TR), which is a major safety concern. TR causes rapid battery heating \cite{xia2023safety}, first at a local level in a few battery cells within a battery module, and then the overheating expands to the module array within a battery pack. Battery packs are installed in EVs; while connected in large numbers, they can form a grid energy storage system. Li-ion battery packs commonly used in EVs have been reported to exhibit high-temperature variations \cite{ramkumar2022review}. Fire accidents due to TR in their Li-ion batteries have been reported for many types of EVs from manufacturers including Tesla (Model S and X), BYD Auto (model e6) \cite{chombo2020review}, and even buses \cite{sun2020review}. \textcolor{blue}{Figure~\ref{fig:abstract}} illustrates the concept of using battery packs, their thermal regulation problems, the pros and cons of cooling strategies, and the thermal management system of the battery proposed in this study. Note that in addition to the TR problem, unbalanced charging occurs in Li-ion batteries subject to a high temperature inhomogeneity due to variations in the temperature-dependent thermophysical properties \cite{habib2023lithium}. This unbalanced charging reduces battery efficiency and reliability. Therefore, both maximum temperature $(\textit{T}_\mathrm{max})$ and temperature uniformity $(\Delta\textit{T})$ are essential factors that need to be addressed when designing thermal management systems for Li-ion batteries.

Battery thermal management systems (BTMS) generally have four cooling methods: air-based, liquid-based, PCM-based, and heat pipes. As an active method, air-based BTMS is generally limited to small electric vehicles \cite{chen2022numerical}. However, the low heat capacity of air can easily result in the inability of air-based BTMS to prevent TR propagation \cite{chen2022numerical} and large temperature inhomogeneities \cite{zhao2021review}. In contrast, heat pipe BTMS have a high cooling capacity that ensures moderate maximum temperatures, but their temperature uniformity is a problem because it can lead to TR \cite{yu2023review}. PCM-based BTMS are widely used passive methods that are suitable for absorbing large amounts of heat generated from rapid charge--discharge cycles. PCMs are excellent local heat absorbers as a result of their large latent heat capacity. However, heat transfer in battery modules with PCM-based cooling is poor because PCMs have low thermal conductivity. After melting, the low heat capacity and poor thermal conductivity of PCMs also lead to large temperature increases \cite{mbulu2021experimental}. Therefore, due to safety concerns caused by poor heat conduction, PCM-based cooling systems are generally not recommended for BTMS \cite{weng2022safety}. Instead, liquid-based BTMS have become the main choice in EVs due to their high heat transfer performance \cite{zhao2018thermal}, albeit there is no latent heat storage that could help mitigate hot spots. To address the issues encountered by single thermal control strategies, hybrid BTMSs have emerged~\cite{rao2016investigation}.

The combination of liquid cooling and mono-PCM has been found to improve the cooling performance of BTMS. Bai et al. \cite{bai2017thermal} reported that for a Li-ion battery module with water channels and PCM in discharge rate of \SI{2}{\crate}, water cooling decreased the maximum temperature, while the PCM contributed to the temperature uniformity of the module. Consequently, Xin et al. \cite{xin2022thermal} analysed the thermal performance of a combined liquid cooling-PCM BTMS at high operating temperatures. Their results showed that high mass flow rates of the coolant guaranteed safe temperature ranges during fast discharging in ambient temperatures as high as \SI{40}{\celsius}. However, a lower coolant mass flow rate is desired to achieve an efficient BTMS because it reduces pumping power. Although effective, high coolant mass flow rates can increase temperature inhomogeneity within battery modules. 

Integration of foam-embedded PCMs with liquid cooling BTMS significantly enhances thermal conductivity, resulting in improvements in cooling performance, particularly for lower coolant velocities. This improvement ensures not only a reduction in maximum temperatures but also better temperature uniformity. Zhao et al. \cite{zhao2023experimental} showed that the use of a copper foam/paraffin-based BTMS in cylindrical Li-ion batteries at a very low discharge rate of \SI{0.5}{\crate} decreased the maximum temperature at high ambient temperatures of up to \SI{45}{\celsius}. However, a high mass flow rate of the coolant increased the inhomogeneity of the temperature field. Another drawback of their proposed battery pack is that only one type of PCM was considered. In contrast, integration of the BTMS with dual PCMs could ensure high cooling efficiencies in a wider range of ambient temperatures. In addition, because of the lower space occupation of the battery pack, prismatic battery cells are commonly used in EVs. Compared to cylindrical batteries, prismatic cells exhibit weak active cooling performance with nonuniformity at high temperature because of a smaller contact surface area with the coolant channels.

Here, we propose and analyse the performance of a multifaceted BTMS comprised of metal foam-embedded dual PCMs with coolant channels. The foam is made of aluminium; the PCMs are 28HC and 35HC \cite{liu2022high}; the coolant is water. This system is designed for high-density rectangular battery cells and is investigated using a computational fluid dynamics (CFD) model, which is verified and validated in this work. To improve the PCM-based cooling systems introduced in the literature, we integrated metal foam into the PCM to increase the thermal conductivity of the PCM elements. In addition, using a dual-PCM configuration, we increased the temperature flexibility of the PCM components. Our proposed foam-embedded dual-PCM BTMS keeps the maximum temperature of the batteries in a safe range. Because the BTMS operates efficiently at low coolant velocities, high-temperature homogeneity is achieved. In fact, the second PCM with a higher melting temperature decreases the battery temperatures during the end of discharge, where the batteries generate excessive heat. In addition, a dual PCM BTMS maintains a lower battery temperature when discharging at high ambient temperatures. In general, the proposed multifaceted BTMS demonstrates good performance in mitigating thermal regulation problems and improving safety.

\section{Proposed multifaceted thermal management}
\begin{figure*}
    \centering
    \includegraphics[width=\textwidth]{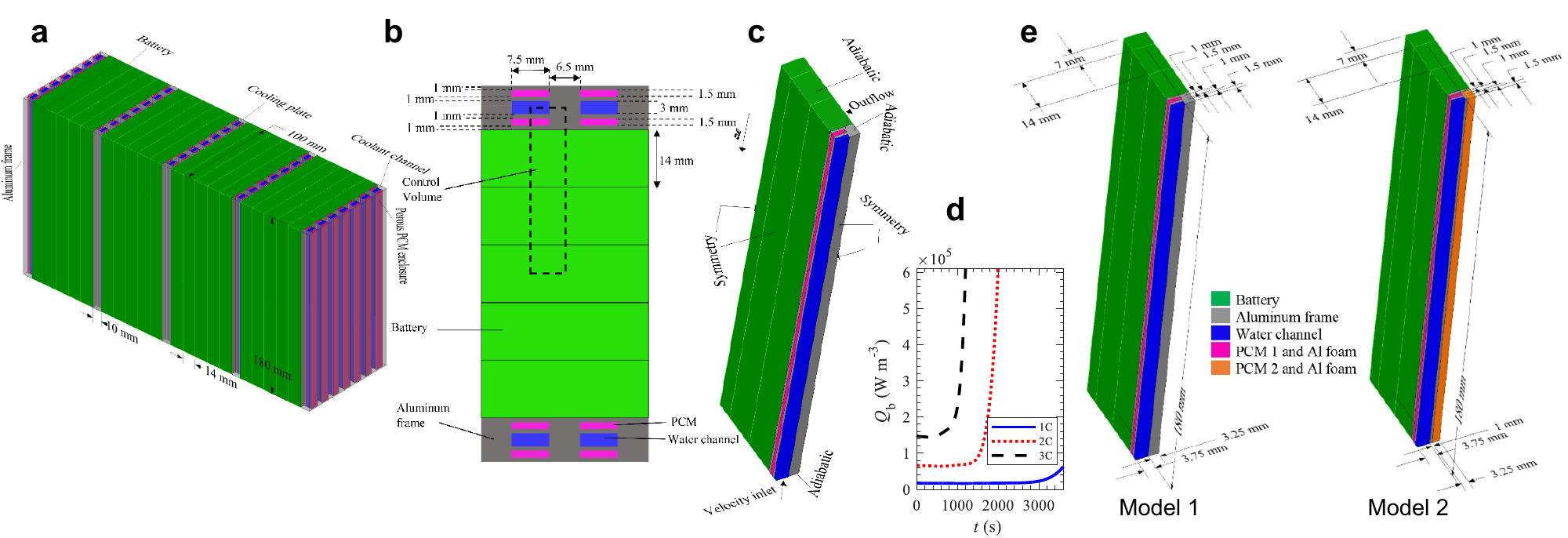}
    \caption{Schematics of the proposed battery packs and integrated battery thermal management system (BTMS). The dimensions and different sections are indicated.
    \textbf{(a)}~Battery pack.
    \textbf{(b)}~Top view of a battery module section.
    \textbf{(c)}~Boundary conditions imposed to the computational fluid dynamics (CFD) model of a battery module.
    \textbf{(d)}~Imposed transient heat generation densities for discharge rates of \SI{1}{\crate}, \SI{2}{\crate} and \SI{3}{\crate}.
    \textbf{(e)}~Two configurations are proposed for the BTMS: single and  double chamber are embodied in Model 1 and Model 2, respectively.}
\label{fig:model}
\end{figure*}

In the battery pack shown in \textcolor{blue}{Fig.~\ref{fig:model}a}, each battery module is composed of five rectangular battery cells. The configuration of BTMS with PCM embedded in metal foam and water channels is illustrated in \textcolor{blue}{Fig.~\ref{fig:model}b}. The upper and lower faces of the battery cells are perfectly insulated. The thermal and physical properties of the battery cell, the aluminium frame, and the coolant are listed in \textcolor{blue}{Table~\ref{tab:properties}}. To achieve an effective multifaceted BTMS, the performance of the proposed active/passive cooling system is analysed for different PCMs whose thermophysical properties are shown in Table~\ref{tab:properties}. The candidate materials for PCMs are selected based on the following criteria: (1) a melting temperature near the operating conditions of the battery, (2) a high amount of latent heat to minimise the volume of the required PCM and create a more compact battery pack, (3) a high value of specific heat to maximise the amount of sensible heat, and (4) cost-effectiveness. In our case, the thermal conductivity of the PCMs is not important since metallic porous fillers are integrated with the PCMs. To analyse the performance of the battery module BTMS, the governing equations for mass, momentum, and energy are solved using a CFD model based on a finite-volume method \cite{Patankar1980}. The boundary conditions imposed on the model are shown in \textcolor{blue}{Fig.~\ref{fig:model} c}. Moreover, for discharge rates of \SI{1}{\crate}, \SI{2}{\crate} and \SI{3}{\crate}, the heat generation of each battery cell during discharge is adopted from \cite{liu2018investigation}, with \textcolor{blue}{Fig.~\ref{fig:model}d} indicating the corresponding transients in volumetric heat generation for each discharge rate. As shown in \textcolor{blue}{Fig.~\ref{fig:model}e}, two BTMS configurations are analysed: a single chamber with two porous PCM containers near each water channel, and double chambers with an extra porous PCM column. The first BTMS model has a water channel with dimensions of $3.75 \times 180 \times \SI{1.5}{\milli\metre}$ and a PCM container of dimensions of $3.75 \times 180 \times \SI{1.5}{\milli\metre}$. The second BTMS model has a water channel with dimensions of $3.75 \times 180 \times \SI{1.5}{\milli\metre}$ and two PCM containers of dimensions of $3.75 \times 180 \times \SI{1.5}{\milli\metre}$ and $3.25 \times 180 \times \SI{4}{\milli\metre}$. These sizes were chosen to fit the different components in the battery module and are depicted in Fig.~\ref{fig:model}b. In addition, the double-chamber BTMS contains two types of PCMs with higher and lower melting points to reduce temperature inhomogeneities and decrease the maximum temperature in the cell.

\begin{table*}
	\caption{Thermophysical properties of water, PCMs, battery and aluminium. The reported properties for the material `battery' are the mass average of representative cell materials. The thermophysical properties include the density $\rho$, the heat capacity $C_\mathrm{p}$, thermal conductivity $\lambda$, dynamic viscosity $\mu$, thermal expansion coefficient $B$, solidus temperature $T_\mathrm{s}$, liquidus temperature $T_\mathrm{l}$ and enthalpy $h$.} \label{tab:properties}
	\begin{tabular}{l c c c c c c c c c }
  \hline
		\textbf{Component} & \textbf{Ref.} & $\rho$ & $C_\mathrm{p}$ & $\lambda$ & $\mu$ & $B$ & $T_\mathrm{s}$ & $T_\mathrm{l}$ & $h$ \\
		&  & (\si{\kilo\gram\per\cubic\meter}) & (\si{\joule\per\kilo\gram\per\kelvin}) &   (\si{\watt\per\meter\per\kelvin}) & (\si{\kilo\gram\meter\per\second}) & (\si{\per\kelvin}) & (\si{\celsius}) & (\si{\celsius}) & (\si{\kilo\joule\per\kilo\gram}) \\
		\hline
		Water	& \cite{haq2015convective} & 998.7 &	4179 &	0.613 &	0.0007 &	0.001	& — & — & — \\
  \hline
		28HC	& \cite{liu2021parametric,liu2013numerical} & 880 (solid)	& 2000	& 0.2 &	0.0269	& 0.00011	& 27	& 29	& 250 \\
		&   & 770 (liquid)		\\	
     \hline
    35HC	& \cite{liu2021parametric,liu2013numerical} & 880 (solid) &	2000 &	0.2 &	0.0269	& 0.00011 &	34	& 36	& 240 \\
	&   &	770 (liquid)	 \\						
     \hline
     44HC &	\cite{liu2021parametric,liu2013numerical} & 800 (solid)	& 2000 &	0.2	& 0.0269 &	0.00011 &	41	& 44 &	250 \\
		& & 700 (liquid)	\\						
    \hline
  n-octadecane	  & \cite{javani2014modeling,abdi2020experimental}  & 814 (solid)	 & 2150 (solid)  &	0.36 (solid)	 & 0.0032	 & 0.00085  &	28  &	30  &	244 \\
	& &	724 (liquid) &	2180 (liquid) &	0.15 (liquid)	\\				
\hline
Battery &	\cite{amalesh2020introducing} & 3000	& 1005.91 & \textit{$\lambda$}$_{x}$ =  \textit{$\lambda$}$_{y}$ = 0.302 &	—	& — &	— &	— &	— \\
	&	&	& & 	\textit{$\lambda$}$_{z}$ = 22.48	 \\				
\hline
Aluminium &	\cite{akbarzadeh2021comparative} & 2719 &	871 &	202.4 &	— &	— &	—	& —	& — \\
\hline
	\end{tabular}
\end{table*}

\section{Methods}
\subsection{Governing equations} 
\textbf{Porous PCM enclosure}. The PCM and aluminium foam are assumed in local thermal non-equilibrium. The metal foam is considered homogeneous and isotropic. For modelling the melting process of the PCM inside the frame filled with aluminium foam, the following mass, momentum and energy  conservation equations were solved

\begin{equation}
    \frac{\mathrm{\partial }\rho }{\mathrm{\partial }t}+\frac{\partial \mathrm{(}\rho u_i\mathrm{)}}{\partial x_i}\mathrm{=0},
        \label{eq1}
    \end{equation}
    
    \vspace{0.25cm} 
    
\begin{equation}
\begin{split}
    \frac{1}{\emptyset }\frac{\partial (\rho u_i)}{\partial t} + \frac{1}{\emptyset^2}\frac{\partial}{\partial x_j}\left(\rho u_i u_j\right) &= -\frac{\partial p}{\partial x_i} \\
    &\begin{aligned}
        &+ \frac{1}{\emptyset }\frac{\partial }{\partial x_j}\left[\mu \left(\frac{\partial u_i}{\partial x_j} + \frac{\partial u_j}{\partial x_i}\right)\right] \\
        &+ \rho \beta \left(T - T_0\right)\overrightarrow{g} - \frac{\mu \overrightarrow{v}}{K} \\
        &- \frac{\rho C_f}{\sqrt{K}}\overrightarrow{v}\left|\overrightarrow{v}\right| + S,
    \end{aligned}
\end{split}
\label{eq2}
\end{equation} 

    \vspace{0.25cm} 

\begin{equation}
    \emptyset \frac{\mathrm{\partial }(\rho H)}{\mathrm{\partial }t}+\emptyset \frac{\partial }{\partial x_j}\left(\rho C_\mathrm{p}u_j T\right)\mathrm{=}\frac{\partial }{\partial x_j}\left({\lambda }_{\mathrm{eff}}\frac{\partial \mathrm{T}}{\partial x_j}\right)+h_{\mathrm{sf}}{\alpha }_{\mathrm{sf}}(T_\mathrm{m}-T),
 \label{eq3}
\end{equation}
where $\emptyset$, $T$, $T_\mathrm{m}$, $h_\mathrm{sf}$ and $\alpha_\mathrm{sf}$ represent the metal foam porosity, PCM temperature, foam temperature, heat transfer coefficient and specific surface area, respectively. Moreover, $K$ and $C_\mathrm{f}$ denote the Al foam permeability and inertial factor and are given as

\begin{equation}
    K=0.00073(1-\emptyset )^{-0.224} d^{-1.11}_\mathrm{f} d^{3.11}_\mathrm{p},
 \label{eq4}
\end{equation}

\begin{equation}
    C_f=0.0012(1-\emptyset)^{-0.132} \left(\frac{d_\mathrm{f}}{d_\mathrm{p}}\right)^{-1.63},
 \label{eq5}
\end{equation}
where $d_\mathrm{p}$ and $d_\mathrm{f}$ represent the pore and ligament diameters of the metal foam and are taken as \SI{1.93}{\milli\meter} and \SI{0.406}{\milli\meter}, respectively \cite{hernandez2005combined}. In addition, $\emptyset$ and PPI of the Al foam are taken as 0.794 and 10, respectively \cite{hernandez2005combined}. For the metal foam, the energy equation is given as

\begin{equation}
    \left(1-\emptyset \right)\frac{\mathrm{\partial }\left({\rho }_{\mathrm{m}}C_{\mathrm{p,m}}T_{\mathrm{m}}\right)}{\mathrm{\partial }t}\mathrm{=}\frac{\partial }{\partial x_j}\left({\lambda }_{\mathrm{m}, \text{eff}}\frac{\partial T_{\mathrm{m}}}{\partial x_j}\right)-h_{\mathrm{sf}}{\alpha }_{\mathrm{sf}}(T_{\mathrm{m}}-T).
 \label{eq6}
\end{equation}

The specific surface area is written as \cite{tian2011numerical}

\begin{equation}
    {\alpha }_{\mathrm{sf}}=\frac{3\pi d_{\mathrm{f}}}{d_{\mathrm{p}}^2}.
 \label{eq7}
\end{equation}

\noindent The heat transfer coefficient is calculated based on the following correlation \cite{zhao2016modeling}

\begin{equation}
    h_{\mathrm{sf}}=\frac{\lambda }{d_\mathrm{f}}\left(0.36+\frac{0.518{Ra}^{0.25}}{{\left(1+{\left(\frac{0.559}{Pr}\right)}^{{9}/{16}}\right)}^{{4}/{9}}}\right),
 \label{eq8}
\end{equation}
where \textit{Pr} and \textit{Ra} are the PCM Prandtl and Rayleigh numbers that are obtained as

\begin{equation}
    Pr=\frac{C_\mathrm{p}\mu}{\lambda},
 \label{eq9}
\end{equation}
\begin{equation}
    Ra=\frac{C_\mathrm{p}{\rho }^2g\beta \pi {d_\mathrm{f}}^3\Delta T}{\lambda \mu}.
 \label{eq10}
\end{equation}

To model the phase-change process of PCMs, we used an enthalpy-porosity method. Consequently, the terms $S$ and $H$ that appeared in Eqs. \eqref{eq2} and \eqref{eq3} represent the momentum sink term and the total enthalpy, respectively. $S$ is given as

\begin{equation}
    S = \frac{{(1 - \gamma)^2}}{{(\gamma^3 + \epsilon)}} A_{\mathrm{mush}} u,
 \label{eq11}
\end{equation}
where $A_\mathrm{mush}$ and $\gamma$ are mushy zone constant and liquid faction, respectively. The details of enthalpy-porosity method can be find in \cite{zhang2022dynamic}. \vspace{2mm}

\textbf{Energy equation in batteries}.
The heat equation for the battery region is based on Fourier's law and is as follows.

\begin{equation}
\begin{split}
    {\rho }_{\mathrm{b}}C_{\mathrm{P,b}}\frac{\mathrm{\partial }T}{\mathrm{\partial }t} &= \frac{\mathrm{\partial }}{\mathrm{\partial }x}\left({\lambda }_{\mathrm{b,x}}\frac{\mathrm{\partial }T}{\mathrm{\partial }x}\right) \\
    &\quad + \frac{\mathrm{\partial }}{\mathrm{\partial }y}\left({\lambda }_{\mathrm{b,y}}\frac{\mathrm{\partial }T}{\mathrm{\partial }y}\right) \\
    &\quad + \frac{\mathrm{\partial }}{\mathrm{\partial }z}\left({\lambda }_{\mathrm{b,z}}\frac{\mathrm{\partial }T}{\mathrm{\partial }z}\right) + Q_\mathrm{b},
\end{split}
\label{eq12}
\end{equation}
where $\lambda_{\mathrm{b},x}$, $\lambda_{\mathrm{b},y}$ and $\lambda_{\mathrm{b},z}$ stand for the battery thermal conductivity in \textit{x}, \textit{y} and \textit{z} directions. In addition, $Q_\mathrm{b}$ is the density of battery heat generation that is adopted from \cite{liu2018investigation} (see Fig.~\ref{fig:model}d). \vspace{2mm}

\textbf{Numerical solution procedure}. The governing equations are solved using a finite volume method in which the velocity-pressure fields are coupled using the SIMPLE algorithm \cite{Patankar1980}. Moreover, an upwind second-order scheme is utilised to discretise the velocity spatial derivatives. In addition, to discretise the spatial derivative of pressure, a standard scheme is used. To satisfy the convergence, the residuals corresponding to the continuity, momentum and energy equations are chosen $10^{-4}$, $10^{-6}$ and $10^{-7}$, respectively. Here, Ansys Fluent software is used to simulate the thermal performance of battery modules.

\subsection{Verification and validation of CFD model}
\begin{figure*}
  \centering
  \includegraphics[width=0.8\linewidth]{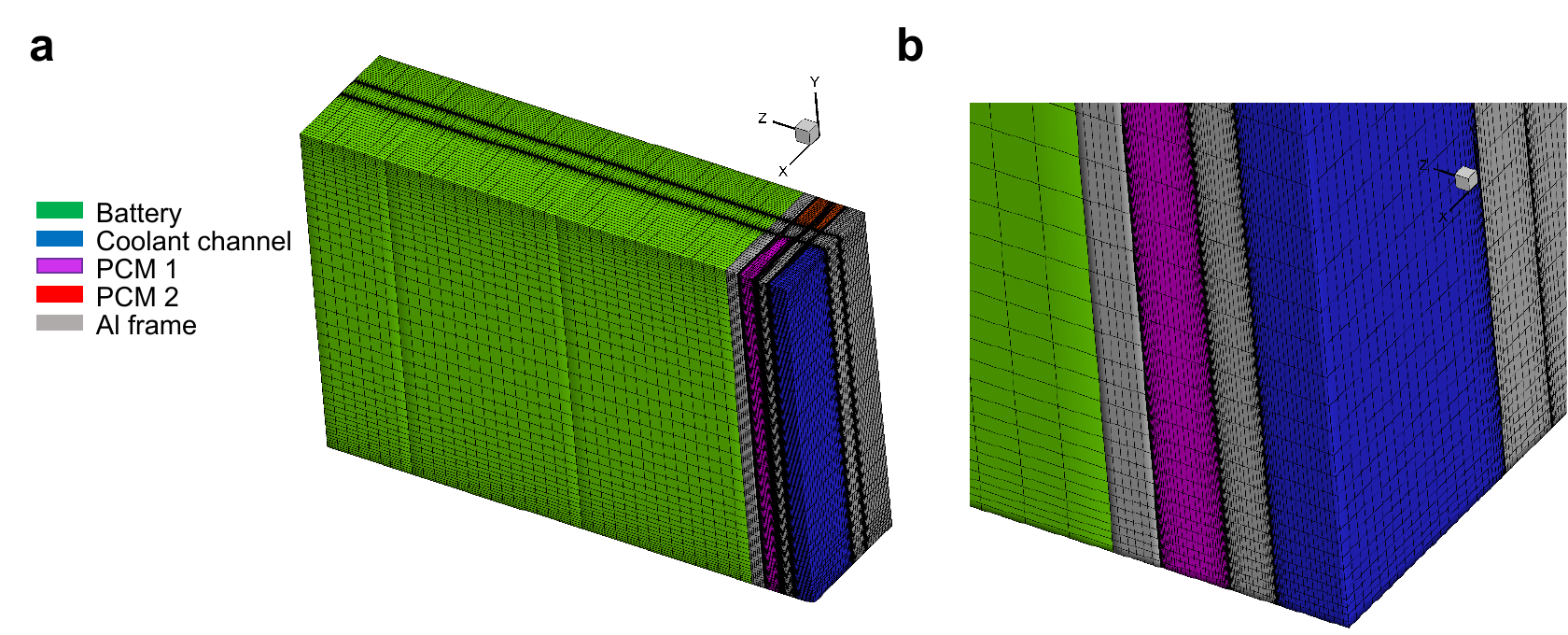}
  \caption[Structured mesh in our computational fluid dynamics (CFD) model]{Structured mesh in our computational fluid dynamics (CFD) model.
  \textbf{(a)}~Grids on the modelled section of a battery module.\textbf{(b)}~A magnified view of grid cells through the BTMS.}
  \label{sifig:Mesh}
\end{figure*}

\textbf{Mesh resolution}.
To guarantee mesh-independent results from our computational fluid dynamics (CFD) model, a mesh refinement study is undertaken with a water-cooled battery thermal management system (BTMS) having a double-chamber battery module (see Fig.~\ref{fig:model}e), integrated with dual PCMs of 28HC and 35HC, as benchmark metric. A water channel mass flow rate of \SI{1.12}{\gram\per\second} ensures active cooling during the discharge process at a discharge rate of \SI{3}{\crate}. The ambient temperature is taken as \SI{25}{\celsius}. The results of the grid independence test are summarised in \textcolor{blue}{Table~\ref{tab:mesh}}. Time-averaged values of the maximum temperature of the batteries, liquid fraction of PCM 1 (first chamber), and liquid fraction of PCM 2 (second chamber), both embedded in aluminium foam,  are compared for five different mesh densities. The hexahedral structural grids shown in \textcolor{blue}{Fig.~\ref{sifig:Mesh}} are generated using Gambit, a commercial meshing software. According to {Table~\ref{tab:mesh}}, a meshed geometry composed of ca. $\sim$600,000 cells (Mesh No. 3) yields precise results with an optimal computational cost. \vspace{2mm}

\begin{table*}
\caption [Grid independence test of the CFD model.] {Mesh independence test of our CFD model. The effect of grid size on time-averaged maximum temperature of batteries and time-averaged liquid fractions of PCM 1 (28 HC) and PCM 2 (35 HC) for a double-chamber battery module at a discharge rate of 3C, initial temperature of \SI{25}{\celsius} and water channel mass flow rate of $\dot m_\mathrm{cha} = \SI{1.12}{\gram\per\second}$.} \label{tab:mesh}

 \centering
  \begin{tabular}{lrcccccc}
    \hline
    \textbf{Mesh No.} & \textbf{Cell No.} & $\bar{T}_\text{max,b}$ (\si{\celsius}) & \textbf{Difference (\%)} & $\bar{\gamma}_\text{PCM 1}$ & \textbf{Difference (\%)} & $\bar{\gamma}_\text{PCM 2}$ & \textbf{Difference (\%)} \\
    \hline
    Mesh 1 & 298,902 & 35.8347 & 1.78 & 0.6412 & 1.25 & 0.0508 & 1.74 \\
    Mesh 2 & 378,048 & 35.5591 & 1.00 & 0.6448 & 0.69 & 0.0513 & 0.77 \\
    Mesh 3 & 597,200 & 35.2418 & 0.09 & 0.6488 & 0.08 & 0.0517 & 0.00 \\
    Mesh 4 & 1,200,544 & 35.2103 & 0.00 & 0.6491 & 0.03 & 0.0517 & 0.00 \\
    Mesh 5 & 2,348,260 & 35.2095 & -- & 0.6493 & -- & 0.0517 & -- \\
    \hline
  \end{tabular}
\end{table*}

\begin{figure}
  \centering
  \includegraphics[width=1\linewidth]{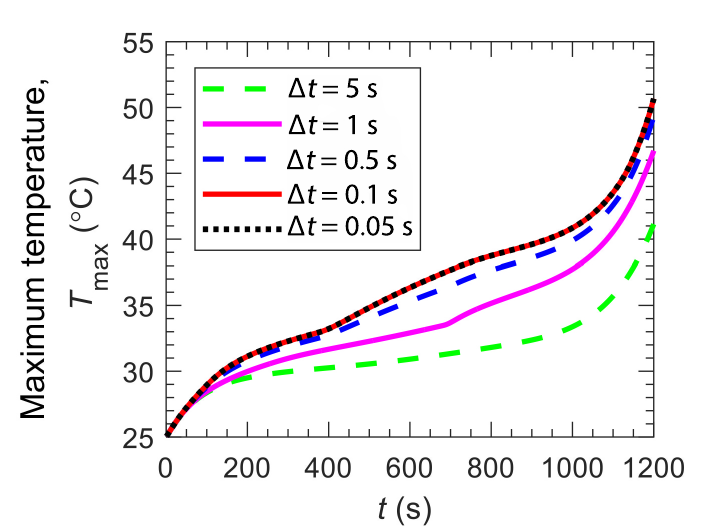}
  \caption[Time independence study for the CFD model] {Time independence study for the CFD model. The effect of time step value (indicated in the legend) on the computed transient maximum temperature in the batteries is presented for a double-chamber BTMS integrated with dual Al foam-embedded PCMs (28HC and 35HC). Simulation conditions include a discharge rate of \SI{3}{\crate}, initial temperature of \SI{25}{\celsius} and water channel mass flow rate of \SI{1.12}{\gram\per\second}.}
  \label{sifig:time_independency}
\end{figure}

\textbf{Time independence study}.
Time step independence tests for our CFD model are carried out for the same metric of a water-cooled, double-chamber battery pack integrated with 28HC and 35HC. The water channel mass flow rate, initial temperature of the battery module and discharge rate are \SI{1.12}{\gram\per\second}, \SI{25}{\celsius} and \SI{3}{\crate}, respectively. According to \textcolor{blue}{Fig.~\ref{sifig:time_independency}}, a time step of \SI{0.1}{\second} yields a precise prediction of the transient maximum temperature in the battery with a maximum error of 0.1\%. This time step used in our CFD modelling. \vspace{2mm}

\begin{figure}
\centering
  \includegraphics[width=0.95\linewidth]{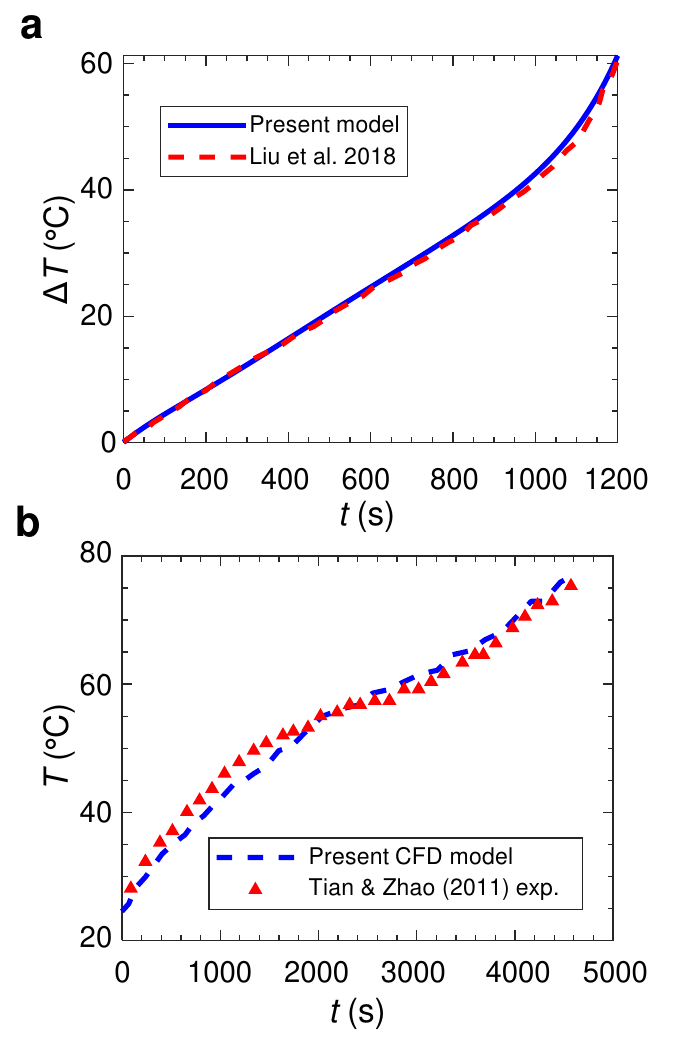}
  \caption[Validation of our CFD model] { Validation of our CFD model.
  \textbf{(a)}~Comparison of the temperature inhomogeneity of batteries obtained by the present model with that presented by Liu et al. \cite{liu2018investigation} for a discharge rate of \SI{3}{\crate} and water mass flow rate of \SI{2.25}{\metre\per\second} for a BTMS with water channels.
  \textbf{(b)} Melting through an aluminum  foam-RT58 PCM composite resulted from the present non-equilibrium model and its comparison with Tian and Zhao’s experimental data \cite{tian2011numerical}.}
  \label{sifig:validation}
\end{figure}

\textbf{Comparison with theoretical study \cite{liu2018investigation}}.
To prove the reliability of the present CFD model, the cooling performance of a rectangular battery pack integrated with water channels is compared with that modelled by Liu et al. \cite{liu2018investigation}. The dimensions of the modelled section of the battery pack are $\SI{6.75}{\milli\meter} \times \SI{180}{\milli\meter} \times \SI{35}{\milli\meter}$. Moreover, the dimensions of the water channel are $\SI{4}{\milli\meter} \times \SI{180}{\milli\meter} \times \SI{1.5}{\milli\meter}$. \textcolor{blue}{Figure~\ref{sifig:validation}a} illustrates the temporal profile of the temperature difference through the battery for the coolant mass flow rate of \SI{2.25}{\gram\per\second} and a discharge rate of \SI{3}{\crate}. The analysis indicates a maximum error of 3.4\% between our CFD results and those reported by Liu et al. \cite{liu2018investigation}. \vspace{2mm}

\textbf{Comparison with experimental study \cite{tian2011numerical}}. To validate our numerical non-equilibrium model applied for heat transfer simulation through PCM–metal foam composites, melting of RT58 within a rectangular enclosure is modelled for comparison with experimental results reported in the literature. The dimensions of the foam-embedded enclosure are $\SI{200}{\milli\meter} \times \SI{120}{\milli\meter} \times \SI{25}{\milli\meter}$. The porosity of the aluminium foam is 0.85. In addition, the bottom surface of the porous enclosure is under a constant heat flux. The temperature profile of the PCM bulk in $y = \SI{8}{\milli\meter}$, where $y$ is the vertical distance from the heating plate, is calculated and compared with the experimental data reported by Tian and Zhao \cite{tian2011numerical}. According to \textcolor{blue}{Fig.~\ref{sifig:validation}b}, there is a good agreement between our results and the experimental data reported by Tian and Zhao \cite{tian2011numerical}. The maximum error is 0.9\% and it was encountered in the initial stages ($\lesssim\SI{1500}{\second}$) of the process, with a better agreement for times exceeding \SI{2000}{\second}.

\section{Results}

\subsection{Selection of BTMS configuration and PCM materials}
We compare various configurations of the BTMS in a battery module, all containing the same water channel. Specifically, three different BTMS configurations are considered: without phase change material (PCM), with a single chamber containing one type of PCM, and a double chamber configuration containing two types of PCM (Fig.~\ref{fig:model}e). The size of the water channel is the same. Additionally, the BTMS of the double chamber configuration is integrated with either one or two types of PCM. We compare different candidate PCMs, namely RT-28HC, RT-35HC, RT-44HC, and n-octadecane. The thermophysical properties of these PCMs are detailed in Table~\ref{tab:properties} with corresponding source in the literature. Furthermore, the water channel mass flow rate and initial battery temperature are set to \SI{2.25}{\gram\per\second} and \SI{25}{\celsius}, respectively. The water mass flow rate per channel is set \SI{2.25}{\gram\per\second} on the basis of achieving a Reynolds number of 611 whereby the flow regime is well within the laminar regime. For example, the critical Reynolds number is about 2000 but our maximum Reynolds number for \SI{2.25}{\gram\per\second} was 611.

\begin{figure*}
    \centering
    \includegraphics[width=1.00\textwidth]{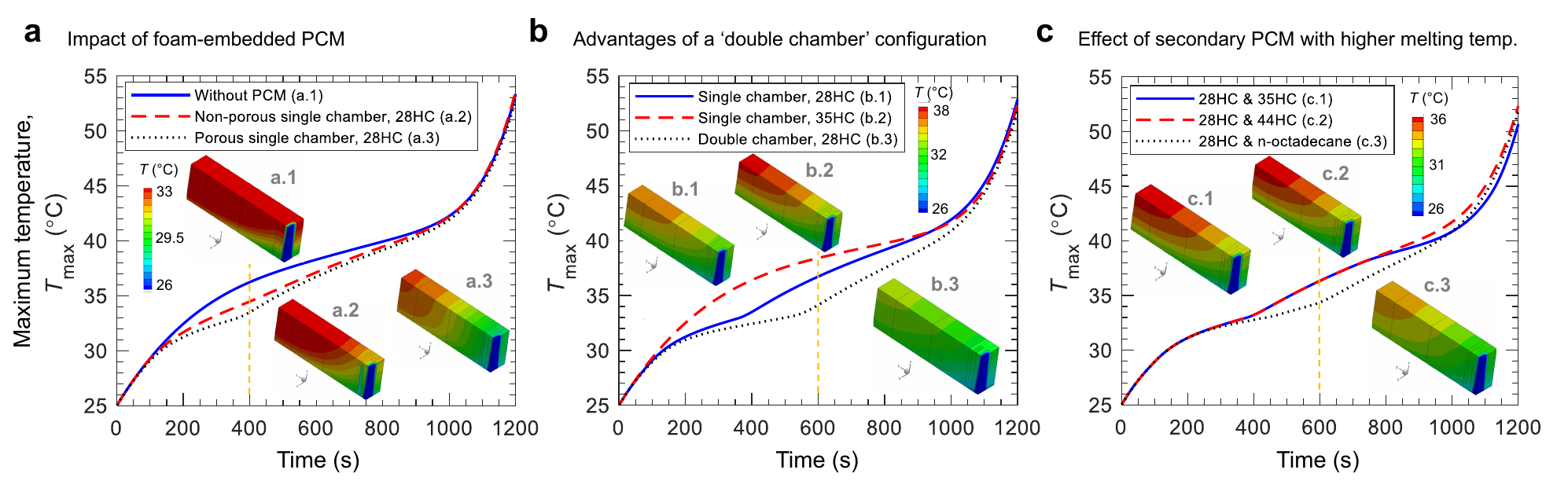}
    \caption{Cooling performance for different BTMS configurations. Discharge rate, channel mass flow rate and ambient temperature are \SI{3}{\crate}, \SI{2.25}{\gram\per\second} and \SI{25}{\celsius}, respectively.
    \textbf{(a)}~Cases without PCM, with non-porous PCM of 28HC and alumnium foam-28HC PCM columns. The insets depict temperature contours on batteries with different BTMS configuration at \SI{400}{\second}.
    \textbf{(b)}~Comparisons between single chamber porous PCM of 28HC, single chamber porous PCM of 35HC and double chambers porous PCM of 28HC in addition to temperature contours of each configuration at $t= \SI{600}{\second}$ depicted in the insets.
    \textbf{(c)}~Comparison of different dual-PCM configurations in dual porous chambers and the corresponding temperature contours at \SI{600}{\second} depicted in the insets. All inset labels are indicated in the corresponding figure legend.}
\label{fig:Configuration}
\end{figure*}

To demonstrate the advantages of inserting a foam-embedded PCM chamber, \textcolor{blue}{Fig.~\ref{fig:Configuration}a} displays the maximum temperature profiles of batteries for three BTMS configurations, i.e. water channel without PCM, with metal-foam-free 28HC PCM, and 28HC PCM/Al foam at a discharge rate of \SI{3}{\crate}. The results show that a battery module integrated with a foam-embedded PCM exhibits a significantly better cooling performance. For instance, at $t = \SI{400}{\second}$, the maximum temperature in the battery module for a case with 28HC PCM/Al foam is \SI{2.7}{\celsius} lower than that of a battery module without PCM and \SI{1.0}{\celsius} lower than that corresponding to a non-porous chamber filled only with 28HC PCM. The temperature distribution contours through the batteries at $t = \SI{400}{\second}$ are depicted in the insets of Fig.~\ref{fig:Configuration}a. However, for operating times higher than approximately \SI{1000}{\second}, the maximum battery temperature becomes approximately the same for all three BTMS cases. The reason is that the entire 28HC PCM melts so, after the latent heat stops playing a role, sensible heat starts accumulating and elevating the temperature of all batteries to similar levels. For example at \SI{1200}{\second}, the $T_\mathrm{max}$ corresponding to Al foam-embedded 28HC PCM is only \SI{0.5}{\celsius} and \SI{0.8}{\celsius} lower than the cases with non-porous 28HC PCM and without PCM, respectively.

We then explore the possibility of introducing 35HC PCM, whose melting point is \SI{7}{\celsius} higher than that of 28HC (see Table~\ref{tab:properties}) and investigate the possibility of increasing the PCM thermal mass with a double-chamber configuration. \textcolor{blue}{Figure~\ref{fig:Configuration}b} compares the thermal performance of three BTMS configurations having a single foam-embedded 28HC PCM, single foam-embedded 35HC PCM, and double-chamber foam-embedded 28HC PCM. These results show that, due to the higher heat storage capacity of the double chamber configuration, the battery temperature is considerably lower than the single chamber case. Additionally, a comparison of single chamber configurations of 28HC and 35HC PCMs reveals that, after $t = \SI{800}{\second}$, 38HC PCM starts to melt and slightly improves the battery cooling performance (i.e. a lower $T_\mathrm{max}$) compared to the case with 28HC. During melting, however, the PCM with lower melting temperature (28HC) performs better. For example, at $t = \SI{600}{\second}$, the $T_\mathrm{max}$ of a battery module with double chamber Al foam/28HC PCM is \SI{2.9}{\celsius} and \SI{4.3}{\celsius} colder than the case with single chambers of 28HC and 35HC, respectively. At $t = \SI{1200}{\second}$, these temperature reductions change to \SI{0.6}{\celsius} and \SI{1.0}{\celsius}, respectively. The insets in Fig.~\ref{fig:Configuration}b show the temperature contours on the three battery modules at $t = \SI{600}{\second}$. A battery configuration with double-chamber foam-embedded 28HC PCM clearly presents and improvement in cooling performance over the single chamber configurations.

We then consider the simultaneous implementation of two types of PCM in a double chamber configuration as means of further improving the effective cooling performance throughout the entire discharging time. \textcolor{blue}{Figure~\ref{fig:Configuration}c} displays $T_\mathrm{max}$ within battery modules with dual PCMs. For $t \lesssim \SI{1000}{\second}$, double porous chambers of 28HC and n-octadecane PCMs maintain the battery at significantly lower temperatures. For example at $t = \SI{600}{\second}$, $T_\mathrm{max}$ of a battery module integrated with foam-embedded 28HC and n-octadecane PCM chambers is \SI{2.0}{\celsius} lower than cases with either 28HC/35HC PCMs or 28HC/44HC. Temperature distributions on battery surfaces for the three different double chamber BTMS configurations are depicted in the insets of Fig.~\ref{fig:Configuration}c. Despite its better cooling performance during melting, the configuration of the dual 28HC/n-octadecane PCMs is unable to efficiently reduce the maximum temperatures of the batteries for operating times of $t \lesssim \SI{1000}{\second}$. To guarantee a reliable cooling process for long-hour discharging conditions, a multifaceted BTMS configuration with 28HC/35HC foam-embedded PCMs is recommended. For example, Fig.~\ref{fig:Configuration}c shows that at $t = \SI{1200}{\second}$ the $T_\mathrm{max}$ of a module configuration with dual 28HC and 35HC foam-embedded PCMs is \SI{1.7}{\celsius} and \SI{1.2}{\celsius} colder than cases with 28HC/44HC and 28HC/n-octadecane PCMs, respectively.

\begin{figure*}
    \centering
    \includegraphics[width=\textwidth]{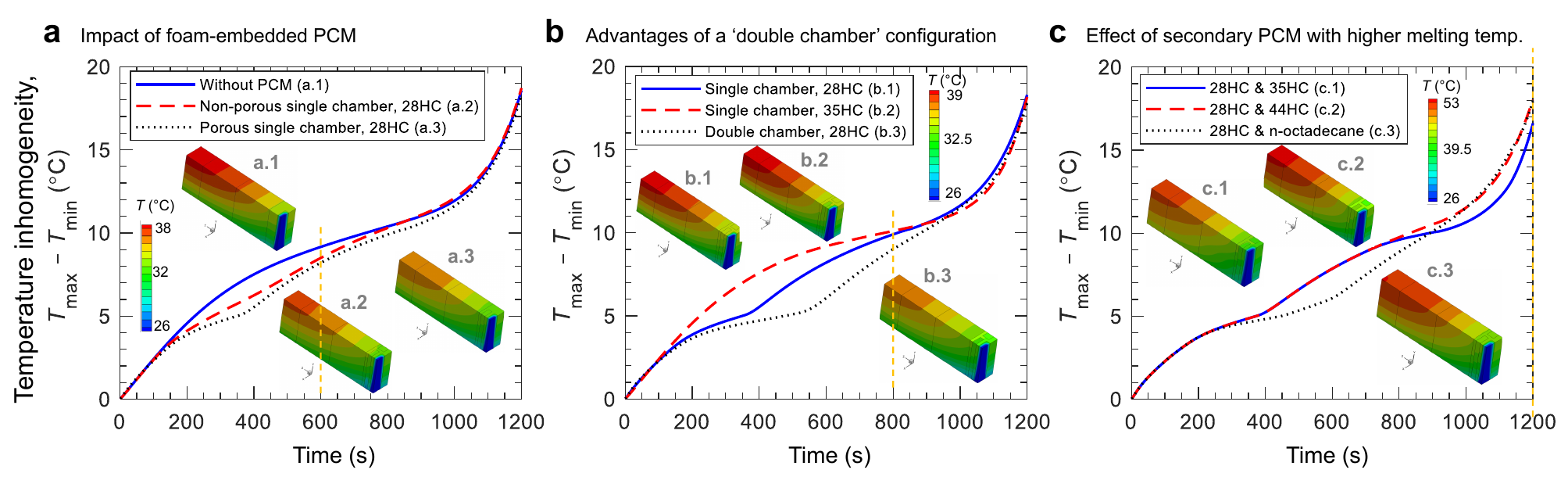}
    \caption{Temperature inhomogeneity of battery modules with different foam-embedded PCM configurations. The results are compared for an operating condition where the discharge rate, ambient temperature and water channel mass flow rate are \SI{3}{\crate}, \SI{25}{\celsius} and \SI{2.25}{\gram\per\second}, respectively.
    \textbf{(a)}~The effect of introducing aluminium foam and 28HC PCM. The insets illustrate the temperature contours at \SI{600}{\second}.
    \textbf{(b)}~Cooling performance of cases with 28HC-Al foam single chamber, 35HC-Al foam single chamber and double chambers porous PCM of 28HC. The temperature distributions on the battery surfaces at \SI{800}{\second} are shown in the insets.
    \textbf{(c)}~The effect of introducing dual PCMs in a double-chamber BTMS. Temperature contours at \textit{t} = \SI{600}{\second} are depicted in the insets. All inset labels are indicated in the corresponding figure legend.}
\label{fig:Configuration2}
\end{figure*}

In addition to reducing the maximum temperature within the battery ($T_\mathrm{max}$), BTMS should also reduce the temperature inhomogeneity ($\Delta T = T_\mathrm{max} - T_\mathrm{min}$). \textcolor{blue}{Figure~\ref{fig:Configuration2}a} shows $\Delta T$ for water cooled BTMS configurations without PCM, foam-free 28HC PCM chamber, and 28HC PCM/Al foam chamber at a discharge rate of \SI{3}{\crate}. The results indicate that a foam-embedded PCM configuration provides better temperature homogeneity than cases without PCM and foam-free PCM. This is an added benefit to that of reducing $T_\mathrm{max}$ in Fig.~\ref{fig:Configuration}a. For example at $t = \SI{400}{\second}$, a battery module equipped with Al foam/28HC PCM is \SI{2.0}{\celsius} and \SI{0.7}{\celsius} more homogeneous (lower $\Delta T$) than those corresponding to cases without PCM and Al foam-free PCM, respectively. Temperature contours for these cases are illustrated at $t = \SI{600}{\second}$ in the insets of Fig.~\ref{fig:Configuration2}a. Further investigation into the double chamber configuration reveals that these are more homogeneous than single chamber configurations during melting, as shown in \textcolor{blue}{Fig.~\ref{fig:Configuration2}b}. At $t \lesssim \SI{950}{\second}$, due to its higher heat removal capacity (higher mass of PCM), a double chamber configuration with 28HC PCM provides reduces temperature inhomogeneity. For example at $t = \SI{600}{\second}$, the $\Delta\textit{T}$ resulting from a double chamber 28HC PCM module is \SI{2.1}{\celsius} and \SI{3.1}{\celsius} lower than the single chamber configurations with 28HC and 35HC PCMs, respectively. Additionally, as shown in \textcolor{blue}{Fig.~\ref{fig:Configuration2}c}, a double chamber configuration with 28HC and 35HC PCMs maintains temperature homogeneity throughout the entire operating time. Comparison of the data in Figs.~\ref{fig:Configuration2}b,c reveals that at $t = \SI{600}{\second}$ and \SI{1200}{\second} the $\Delta T$ corresponding to a double chamber configuration with 28HC and 35HC PCMs are \SI{1.8}{\celsius} and \SI{1.2}{\celsius} lower than those for a battery module with a double chamber 28HC PCM.

\subsection{Performance of BTMS with foam-embedded PCMs}  

\begin{figure*}
    \centering
    \includegraphics[width=\textwidth]{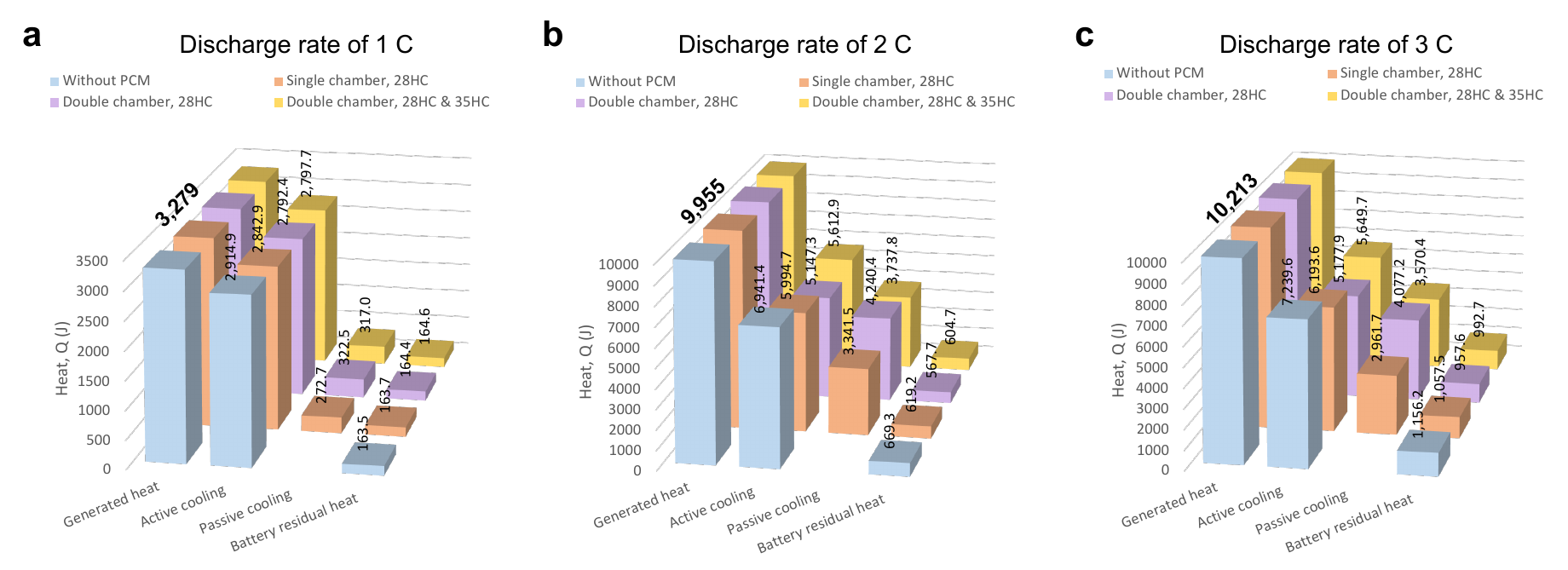}
    \caption{Heat removal breakdown for different battery module configurations. The portions of active cooling (water channel), passive cooling (porous PCM) and residual heat within the batteries for a section of the battery module shown in (Fig. ~\ref{fig:model}e) are shown in an ambient temperature of 25$^{\circ}$C. The discharge rate is
    \textbf{(a)}~\SI{1}{\crate},
    \textbf{(b)}~\SI{2}{\crate}, and
    \textbf{(c)}~\SI{3}{\crate}.}
\label{fig:qportion}
\end{figure*}

We compared the cooling performance of actively cooled battery modules without PCM (water channels only) with concurrent active / passive cooling approaches that have PCMs for discharge rates of \SI{1}{\crate}, \SI{2}{\crate}, and \SI{3}{\crate}. Four different water-cooled BTMS configurations were considered: (1)~baseline case without PCM, (2)~single chamber with Al foam/28HC PCM, (3)~double chambers with Al foam/28HC PCM, and (4)~double chambers with Al foam / dual PCM of 28HC and 35HC PCM. The water mass flow rate and initial temperature were set to \SI{2.25}{\gram\per\second} and \SI{25}{\celsius}, respectively. \textcolor{blue}{Figure~\ref{fig:qportion}} illustrates the breakdown of the heat generated and removed for the discharge rates of \SI{1}{\crate}, \SI{2}{\crate}, and \SI{3}{\crate}. According to Fig.~\ref{fig:qportion}a for a discharge rate of \SI{1}{\crate}, most of the total heat generated by the batteries is actively removed by the water channels for cases without PCM (88.9\% of the total heat generated), single chamber 28HC (86. 7\%), double chambers 28HC (85. 2\%) and double chambers 28HC/35HC (85. 3\%). Therefore, the integration of metal foam-embedded PCMs with a battery module at a discharge rate of \SI{1}{\crate} may not be economically viable. However, for discharge rates of \SI{2}{\crate} and \SI{3}{\crate}, the role of foam-embedded PCMs in the cooling process becomes more important. For example, for discharge rates of \SI{2}{\crate} and \SI{3}{\crate} in Fig.~\ref{fig:qportion}b and Fig.~\ref{fig:qportion}c, respectively, 37.5\% and 35.0\% of the battery heat is removed by the PCMs in the fourth multifaceted configuration of a double-chamber foam-embedded 28HC/35HC. For a double chamber mono-pcm BTMS, these values are 42.6\% and 39.9\% for discharge rates of \SI{2}{\crate} and \SI{3}{\crate}, respectively. Although the amount of heat removed by PCMs in the double chamber 28HC PCM configuration is slightly higher than that dissipated by PCMs in the 28HC/35HC configuration, the latter provides lower $T_\mathrm{max}$ and $\Delta T$ for long-period operations at discharge rates \SI{2}{\crate} and \SI{3}{\crate} discharge rates (recall Figs.~\ref{fig:Configuration}c and \ref{fig:Configuration2}c).

\begin{figure*}
    \centering
    \includegraphics[width=\textwidth]{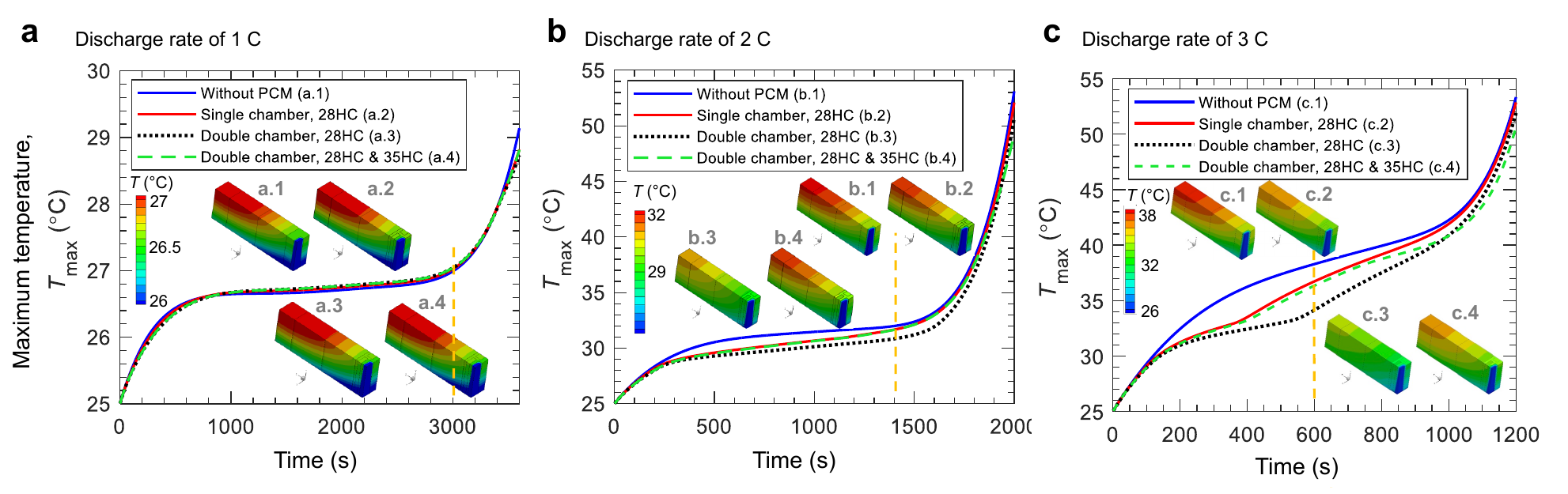}
    \caption{Cooling performance of four battery modules at different discharge rates. The cooling performance is presented in terms of the maximum temperature in the cell, $T_\mathrm{max}$ for four battery modules: without PCM, single-chamber (porous PCM) and double-chamber configurations with 28HC PCM and 28HC/35HC PCM. The ambient temperature and the coolant mass flow rate per channel are taken as \SI{25}{\celsius} and \SI{2.25}{\gram\per\second}, respectively. 
    \textbf{(a)}~Discharge rate of \SI{1}{\crate}; insets depict the temperature contours of the batteries at $t = \SI{3000}{\second}$.
    \textbf{(b)}~Discharge rate of \SI{2}{\crate}; insets show the temperature contours at $t = \SI{1400}{\second}$.
    \textbf{(c)}~Discharge rate of \SI{3}{\crate}; temperature contours at $t = \SI{600}{\second}$ are shown in the insets. All inset labels are indicated in the corresponding figure legend.
    }
\label{fig:discharge_rates}
\end{figure*}

According to the results shown in \textcolor{blue}{Fig.~\ref{fig:discharge_rates}a} for a discharge rate of \SI{1}{\crate}, the integration of porous PCMs in the BTMS does not improve its cooling performance for discharge times of $t<\SI{3400}{\second}$. However, in harsh ambient temperatures, a 25HC/35HC double chamber configuration can keep the batteries at lower temperatures (discussed later in Sec. \ref{sec:ambtemp}). Regarding \textcolor{blue}{Fig.~\ref{fig:discharge_rates}b}, in an operating condition where the discharge rate is \SI{2}{\crate},the cooling performance of a battery module with double chambers filled with 28HC/35HC PCM becomes better than the double chamber 28HC PCM configuration at $t \gtrsim\SI{1900}{\second}$. For example at $t = \SI{2000}{\second}$, $T_\mathrm{max}$ and $\Delta T$ of the battery module of double chamber 28HC/35HC are \SI{1.3}{\celsius} and \SI{2.3}{\celsius} lower than those of double chamber 28HC (see \textcolor{blue} {Appendix A1} for corresponding $\Delta T$ and PCM liquid fraction $\gamma$). As shown in \textcolor{blue}{Fig.~\ref{fig:discharge_rates}c}, when the battery operates at a discharge rate of \SI{3}{\crate}, for $t\lesssim\SI{1000}{\second}$, the double chamber Al foam / 28HC PCM system provides the best cooling performance (and temperature homogeneity) in the batteries. For $t \gtrsim \SI{1000}{\second}$, a configuration of double chambers with Al foam/dual 28HC and 35HC PCMs guarantees the minimum $\Delta T$ and thermal stress (Fig.~S?). For example at $t = \SI{1200}{\second}$, for 28HC/35HC PCMs with double chamber foam embedded in discharge rate of \SI{3}{\crate}, $T_\mathrm{max}$ and $\Delta T$ are \SI{1.3}{\celsius} and \SI{1.2}{\celsius} lower than the case of 28HC PCMs with double chamber foam embedded.

\begin{figure*}
    \centering
    \includegraphics[width=0.9\textwidth]{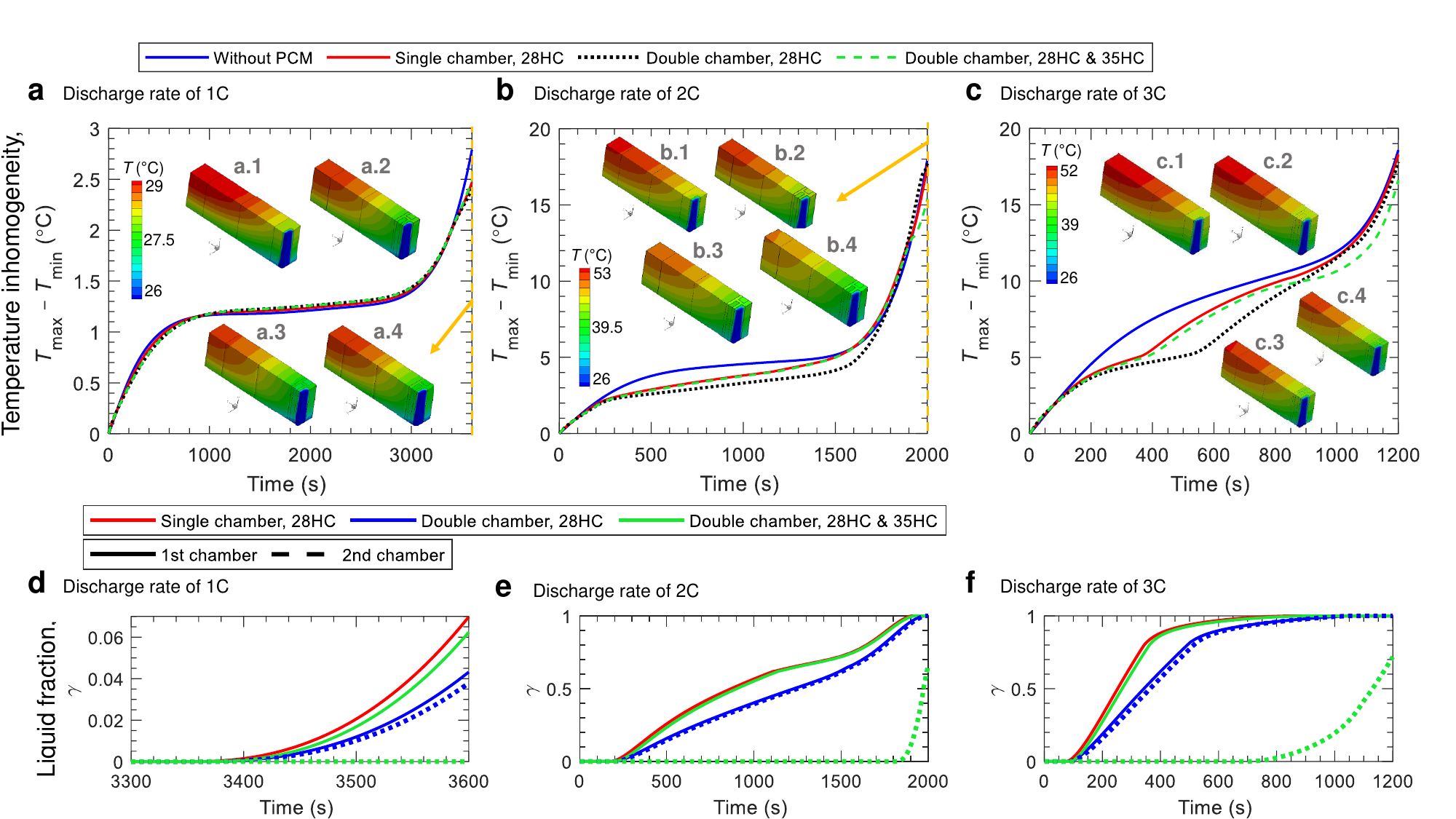}
    \caption[Temperature inhomogeneity and PCM liquid fraction for different BTMS approaches]{Cooling performance of four battery modules at different discharge rates.Cooling performance is presented in terms of the temperature inhomogeneity in the cell, $\Delta T = T_\mathrm{max}-T_\mathrm{min}$ for four battery modules: without PCM (a.1, b.1 and c.1), single chamber porous PCM (a.2, b.2 and c.2) and double chamber configurations with 28HC PCM (a.3, b.3 and c.3) and 28HC/35HC PCMs (a.4, b.4 and c.4). The insets show the temperature profile on the cell surfaces. The ambient temperature and the coolant mass flow rate per channel were \SI{25}{\celsius} and \SI{2.25}{\gram\per\second}, respectively. 
    \textbf{(a)}~Discharge rate of \SI{1}{\crate}; insets depict the temperature contours of the batteries at $t = \SI{3600}{\second}$.
    \textbf{(b)}~Discharge rate of \SI{2}{\crate}; insets show the temperature contours at $t = \SI{2000}{\second}$.
    \textbf{(c)}~Discharge rate of \SI{3}{\crate}; temperature contours at $t = \SI{1200}{\second}$ are shown in the insets.
    \textbf{(d)--(f)}~Corresponding PCM liquid fraction for each discharge rate analysed in this study.
    }
\label{fig:discharge_rates2}
\end{figure*}

\textcolor{blue}{Figures~\ref{fig:discharge_rates2}a-c} show the inhomogeneity of the battery temperature of four different water-cooled BTMS configurations: (1)~baseline case without PCM, (2)~double chambers with Al foam/28HC PCM, (3)~double chambers with Al foam/28HC PCM, and (4)~double chambers with Al foam / dual PCM of 28HC and 35HC PCM discharge rates of \SI{1}{\crate}, \SI{2}{\crate}, and \SI{3}{\crate}. The water channel mass flow rate and initial temperature were set to \SI{2.25}{\gram\per\second} and \SI{25}{\celsius}, respectively. According to Fig.~\ref{fig:discharge_rates2}c, at a discharge rate of \SI{3}{\crate}, a dual-PCM BTMS reduces $\Delta T$ up to \SI{2.0}{\celsius} than the case without PCM and \SI{1.2}{\celsius} compared to a double chamber mono-PCM BTMS. The transient fraction of the PCM liquid at different discharge rates is shown in \textcolor{blue}{Figs.~\ref{fig:discharge_rates2}d-f}.
As an important point, it should be mentioned that installing foam-PCM columns in between batteries, in addition to the cooling plate, significantly improves the cooling performance of the battery module (see the Appendix). However, when PCM components are used between the batteries, the latent heat released during the off-use cycle, i.e. standby, could be directly absorbed by the adjacent batteries. This could lead to thermal runaway problems. Therefore, in the present work, the foam PCM elements are integrated only with the cooling plate, not between the batteries.

\subsection{Effect of coolant flow rate on BTMS performance}
\begin{figure*}
    \centering
    \includegraphics[width=\textwidth]{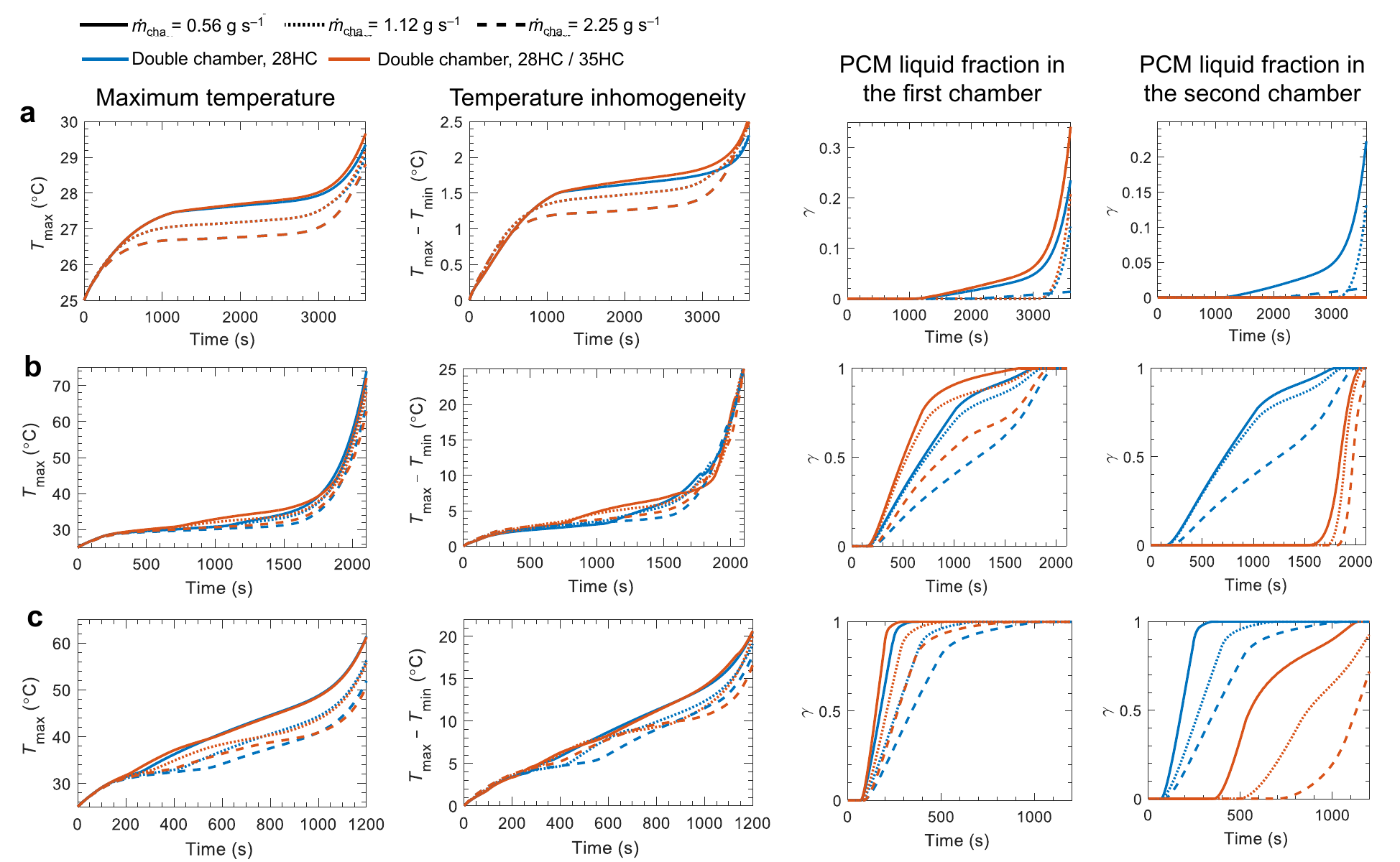}
    \caption{Effect of channel mass flow rate on the cooling performance of BTMS. Water-cooled BTMS include double-chamber configurations with single- and dual-PCMs.
    The ambient temperature is \SI{25}{\celsius} for all cases.
    The peak temperature in the battery ($T_\mathrm{max}$), the inhomogeneity of the temperature within the battery ($T_\mathrm{max}-T_\mathrm{min}$), and liquid fractions ($\gamma$) of the PCM inside the first and second chambers are shown as a function of time.
    Analyses were carried out for discharge rates of
    \textbf{(a)}~\SI{1}{\crate}, 
    \textbf{(b)}~\SI{2}{\crate}, and
    \textbf{(c)}~\SI{3}{\crate}.
    All plots share the same legend, indicated on the top left. Three water channel mass flow rates $\dot m_\mathrm{cha}$ were analysed for a double-chamber configuration containing only 28HC and both 28HC and 35HC.}
\label{fig:massflow}
\end{figure*}

We evaluated the cooling performance of double-chamber battery modules for the cases with only 28HC PCM and dual-PCM 28HC/35HC with water channel mass flow rates of $\dot m_\mathrm{cha} = \SI{0.56}{\gram\per\second}$, \SI{1.12}{\gram\per\second}, and \SI{2.25}{\gram\per\second} for ambient temperature of \SI{25}{\celsius}. This mass flow rate $\dot m_\mathrm{cha}$ is for a single channel, noting that the proposed BTMS module has 14 channels. \textcolor{blue}{Figure~\ref{fig:massflow}a} shows that, for a discharge rate of \SI{1}{\crate}, the cooling performance of the 28HC double chamber PCM becomes slightly more efficient after \SI{1200}{\second}, \SI{3150}{\second}, and \SI{3400}{\second} for water channel mass flow rates of \SI{0.56}{\gram\per\second}, \SI{1.12}{\gram\per\second}, and \SI{2.25}{\gram\per\second}, respectively. Before the aforementioned critical times, the cooling performance of the two configurations is the same. For example, at a water channel mass flow rate of \SI{0.56}{\gram\per\second} at $t = \SI{3600}{\second}$, $T_\mathrm{max}$ and $\Delta T$ corresponding to the double chamber configuration of the 28HC PCM are \SI{0.3}{\celsius} lower than the case of 28HC/35HC PCMs.

For the discharge rate of \SI{2}{\crate}, the cooling performance of double chamber battery modules with 28HC PCM and 28HC/35HC PCMs is depicted in \textcolor{blue}{Fig.~\ref{fig:massflow}b}. As can be inferred from the figure, for all mass flow rates and operating times $t<\SI{1760}{\second}$, the single 28HC PCM results in lower $T_\mathrm{max}$ and $\Delta T$. In contrast, as the discharging progresses further, the dual-28HC/35HC-PCM configuration shows a better thermal performance as evidenced by its lower $T_\mathrm{max}$. For example, for \SI{2.25}{\gram\per\second} at $t = \SI{2100}{\second}$, the reductions (from the case of 28HC) in $T_\mathrm{max}$ and $\Delta\textit{T}$ for the configuration of 28HC/35HC PCM are \SI{1.9}{\celsius} and \SI{0.6}{\celsius} for a mass flow rate of the water channel of \SI{0.56}{\gram\per\second}, while these reductions become \SI{0.5}{\celsius} and \SI{0.4}{\celsius} for a mass flow rate of \SI{2.25}{\gram\per\second}.

For a higher discharge rate of \SI{3}{\crate}, \textcolor{blue}{Fig.~\ref{fig:massflow}c} depicts the cooling performance of the proposed BTMS. The time at which the `best' configuration (with lowest $T_\mathrm{max}$) changes from single (28HC) to dual-PCM (28HC/35HC) for mass flow rates of \SI{0.56}{\gram\per\second}, \SI{1.12}{\gram\per\second}, and \SI{2.25}{\gram\per\second} is \SI{536}{\second}, \SI{765}{\second} and \SI{990}{\second}, respectively. This suggests that BTMS with dual-PCM are less effective short term, and for low channel mass flow rates where convective heat transfer is less dominant. In contrast, for long operating times and high mass flow rates, the cooling performance of the dual 28HC/35HC configuration is superior. For example, at $t = \SI{1200}{\second}$, the reductions in $T_\mathrm{max}$ and $\Delta T$ for a battery module with 28HC/35HC PCMs compared to a double chamber 28HC configuration are \SI{0.2}{\celsius} and \SI{0.1}{\celsius},respectively, for a channel mass flow rate of \SI{0.56}{\gram\per\second}. In contrast, for a mass flow rate of \SI{2.25}{\gram\per\second}, these reductions in $T_\mathrm{max}$ and $\Delta T$ become \SI{1.3}{\celsius} and \SI{1.2}{\celsius}, respectively. As a result, the usage of multi PCMs of 28HC and 35HC is an effective passive method to reduce the maximum temperature of the battery and increase temperature homogeneity. Furthermore, as can be seen from \textcolor{blue}{Fig.~\ref{fig:massflow}c}, after $t = \SI{1200}{\second}$, 7.5\% and 26.9\% of the initial volume of 35HC PCM is not melted at flow rates of \SI{1.12}{\gram\per\second} and \SI{2.25}{\gram\per\second}, respectively. Hence, if the discharging process continues further, chamber 2 (35HC) will play a role in the temperature regulation of the battery pack. In the mono-PCM (28HC) configuration, all PCM in both chambers are melted before $t = \SI{1200}{\second}$ at the three water flow rates.   

\subsection{Effect of ambient temperatures on BTMS performance}\label{sec:ambtemp}
\begin{figure*}
    \centering
    \includegraphics[width=0.8\textwidth]{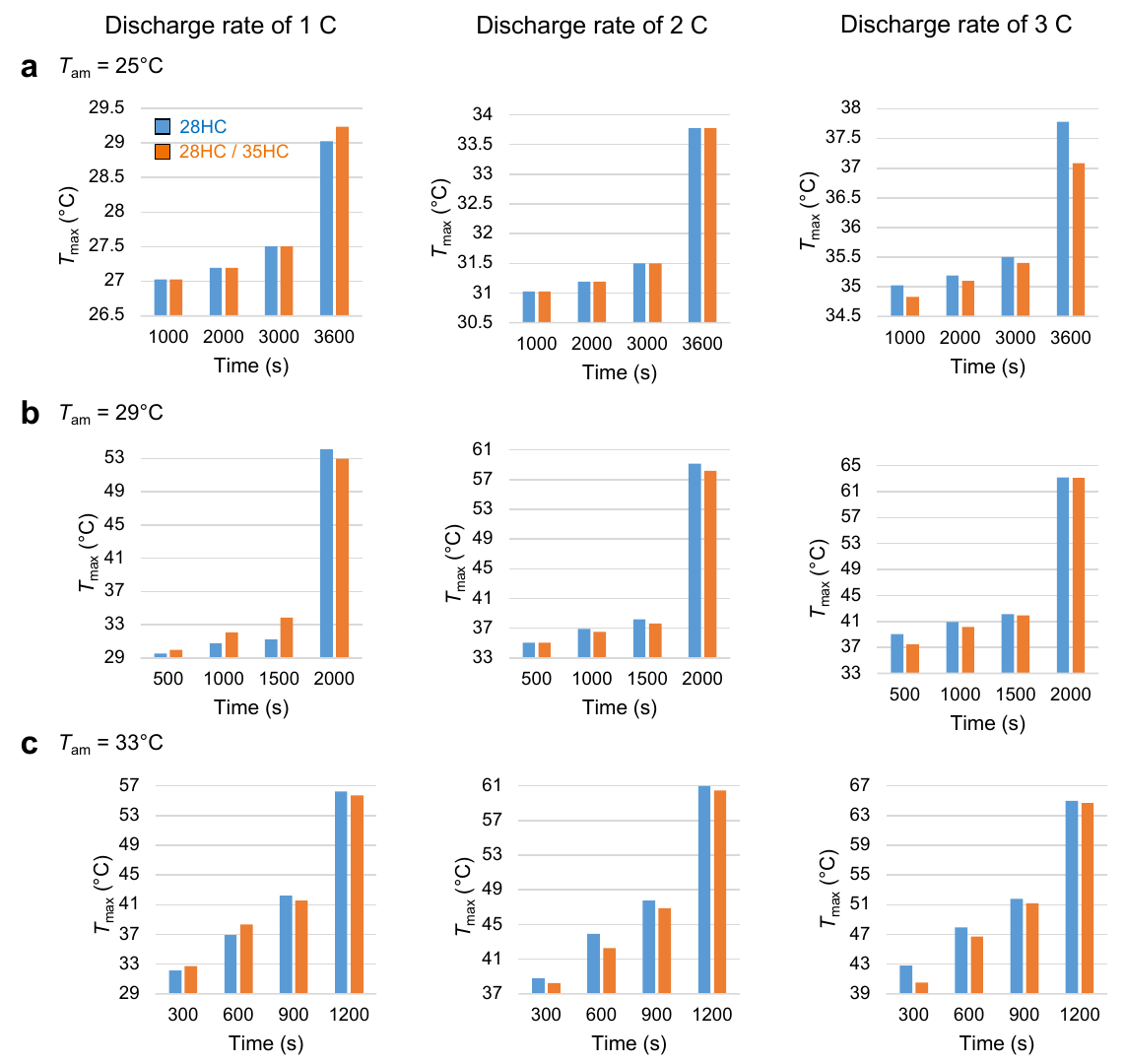}
    \caption{Effect of ambient temperature on the cooling performance of BTMS.
    The analyses were performed for a water-cooled BTMS with double-chamber configuration. 28HC fills both chambers (blue bars) or 28HC and 35HC fill chambers 1 and 2, respectively (orange bars). The water channel mass flow rate was set to \SI{1.12}{\gram\per\second}. The maximum battery temperature $T_\mathrm{max}$ is reported for ambient temperatures of \textbf{(a)}~\SI{25}{\celsius}, \textbf{(b)}~\SI{29}{\celsius}, and \textbf{(c)}~\SI{33}{\celsius}.}
\label{fig:temperature}.
\end{figure*}

To gain insight into the cooling performance of double chamber battery module configurations with either 28HC or 28HC/35HC under high ambient temperatures, simulations are conducted for ambient temperatures $T_\mathrm{am}$ representative of hot summer days. The ambient temperatures analysed were $T_\mathrm{am} = \SI{25}{\celsius}$ (`comfortable ambient temperature), \SI{29}{\celsius} and \SI{33}{\celsius} (hot summer day) with a channel mass flow rate of \SI{1.12}{\gram\per\second}. The results for discharge rates of \SI{1}{\crate}, \SI{2}{\crate} and \SI{3}{\crate} are shown in \textcolor{blue}{Fig.~\ref{fig:temperature}}. We found that the 28HC/35HC dual-PCM configuration significantly improves the safety of the battery module at an ambient temperature of \SI{33}{\celsius}. For example, for a discharge rate of \SI{1}{\crate} at $t = \SI{3600}{\second}$, $T_\mathrm{max}$ is \SI{0.7}{\celsius} lower in 28HC/35HC compared to 28HC cases. Regarding the discharge rate of \SI{2}{\crate}, the values of $T_\mathrm{max}$ for the dual-28HC/35HC-PCM configuration are lower at $t > \SI{2000}{\second}$ for $T_\mathrm{am} = \SI{25}{\celsius}$. Furthermore, dual-28HC/35HC-PCM yields a lower temperature for $t>\SI{1000}{\second}$ and $T_\mathrm{am} = \SI{29}{\celsius}$, and at $t > \SI{500}{\second}$ and $T_\mathrm{am} = \SI{33}{\celsius}$. When the battery is at a discharge rate \SI{3}{\crate}, the threshold times at which the maximum battery temperature of the 28HC/35HC configuration is lower than the case of the double chamber with 28HC PCM are approximately \SI{900}{\second} for $T_\mathrm{am} = \SI{25}{\celsius}$ and approximately \SI{300}{\second} for $T_\mathrm{am} = \SI{29}{\celsius}$ and \SI{33}{\celsius}.

\section{Conclusions}
The maximum temperature and inhomogeneity of the temperature within the lithium ion battery packs determine their degradation rates and the risk of failure. When batteries operate at high ambient temperatures, cell temperatures could potentially reach critical levels, particularly at discharge rates exceeding \SI{1}{\crate}. Thermal regulation problems lead to safety concerns and motivate further research. Here, we proposed a multifaceted battery thermal management system (BTMS) design that integrates active liquid cooling (through water channels) with latent heat thermal removal (with foam-embedded phase-change materials, or PCMs). The effects of implementing two types of PCM in the passive section of the BTMS were investigated, with one PCM having a low melting temperature (i.e. 28HC at ca. \SI{28}{\celsius}) and another one a high melting temperature (i.e. 35HC at ca. \SI{35}{\celsius}). We found that in the proposed multifaceted dual PCM BTMS, 28HC effectively absorbed the heat generated during the initial stages of the battery discharging process, while 35HC reduced the maximum temperature caused by excessive heat generation toward the end of the discharging phase. We analysed the performance of the proposed cooling system under different operating conditions to determine cases where the cooling effectiveness of the dual-PCM 28HC/35HC configuration exceeds that of the double chamber 28HC BTMS. We demonstrated that a multifaceted passive/active BTMS with dual-PCM configuration is an effective means for mitigating thermal runaway or other degradation issues brought about by thermal inhomogeneities.

We numerically investigated three materials as secondary PCM with a higher melting temperature: 35HC, 44HC, and n-octadecane, while 28HC was used as the primary PCM. We found that the dual-PCM 28HC/35HC BTMS provides temperatures up to \SI{1.7}{\celsius} and \SI{1.2}{\celsius} lower than those cases with 28HC/44HC and 28HC/n-octadecane PCMs, respectively, for an ambient temperature of $T_\mathrm{am} = \SI{25}{\celsius}$ and a water channel mass flow rate of $\dot m_\mathrm{cha} = \SI{2.25}{\gram\per\second}$. For the same operating conditions and the case in which both chambers are filled with Al foam/28HC PCM, 42.6\% and 39.9\% of the heat is removed by the foam-embedded PCM for discharge rates of \SI{2}{\crate} and \SI{3}{\crate}, respectively. These values become 37. 5\% and 35. 0\% for the double chamber configuration 28HC / 35HC. Our findings suggest that dual-PCM designs provide considerably lower maximum temperature while reducing temperature inhomogeneity. For example, the values of $T_\mathrm{max}$ and $\Delta T$ with a 28HC/35HC BTMS are \SI{1.8}{\celsius} and \SI{1.3}{\celsius} lower than those of the double chamber 28HC for discharge rates of \SI{2}{\crate} and \SI{3}{\crate}, respectively.

For a discharge rate of \SI{1}{\crate} and a channel mass flow rate of $\dot m_\mathrm{cha} = \SI{2.25}{\gram\per\second}$, the cooling performances of the double-chamber and single-chamber configurations are nearly identical for an ambient temperature of $T_\mathrm{am} = \SI{25}{\celsius}$. Generally, for battery operations at high ambient temperatures, the 28HC / 35HC double chamber results in lower maximum battery temperatures. For example, in the case where $\dot m_\mathrm{cha} = \SI{1.12}{\gram\per\second}$ and $T_\mathrm{am} = \SI{33}{\celsius}$, the maximum temperature corresponding to a double chamber 28HC / 35HC configuration is up to \SI{2.0}{\celsius} lower than those of a double chamber packed only with 28HC PCM. However, a lower coolant mass flow rate could be applied for a double chamber 28HC configuration. This may have benefits in reducing the pumping power, albeit the low flow rates in our BTMS suggest that pumping power is not a major concern. The experimental implementation and energy efficiency analysis of the proposed BTMS could be the focus of future studies.

\setlength{\nomlabelwidth}{1.2cm}
\setlength{\nomitemsep}{-\parsep}
\nomenclature[A, 01]{\(\mathbf{\mathit{g}}\)}{Acceleration due to gravity}
\nomenclature[A, 02]{\(T_{\mathrm{max}}\)}{Maximum battery temperature}
\nomenclature[A, 03]{$\Delta\textit{T}$}{Temperature inhomogeneity}
\nomenclature[A, 04]{$u_i$}{Velocity in \textit{i}-direction}
\nomenclature[A, 05]{$\textit{P}$}{Pressure}
\nomenclature[A, 06]{\(\mathrm{CFD}\)}{Computational fluid dynamics}
\nomenclature[A, 07]{$h_\mathrm{sf}$}{Heat transfer coefficient}
\nomenclature[A, 08]{$d_\mathrm{p}$}{The pore diameter of the foam}
\nomenclature[A, 9]{$d_\mathrm{f}$}{Ligament diameter of the metal foam}
\nomenclature[A, 10]{$C_{\mathrm{p}}$}{Specific heat}
\nomenclature[A, 11]{$k$}{Al foam permeability}
\nomenclature[A, 12]{$C_\mathrm{f}$}{Inertial factor of metal foam}
\nomenclature[A, 13]{$Pr$}{Prandtl number}
\nomenclature[A, 14]{$Ra$}{Rayleigh number}
\nomenclature[A, 15]{$\textit{S}$}{Momentum sink term}
\nomenclature[A, 16]{$\textit{h}$}{Enthalpy of fussion}
\nomenclature[A, 17]{$\textit{H}$}{Total enthalpy}
\nomenclature[A, 18]{$A_\mathrm{mush}$}{Mushy zone constant}
\nomenclature[A, 19]{ $T_\mathrm{s}$}{Solidus temperature}
\nomenclature[A, 20]{ $T_\mathrm{l}$}{Liquidus temperature}
\nomenclature[A, 21]{  $Q_\mathrm{b}$}{Battery heat generation density}
\nomenclature[A, 22]{  $\dot m$}{Mass flow rate}
\nomenclature[A, 23]{  $Q$}{Heat}
\nomenclature[A, 24]{\(T_{\mathrm{m}}\)}{Temperature of the metal foam}

\nomenclature[G, 01]{$\rho$}{Density}
\nomenclature[G, 02]{$\emptyset$}{Porosity of the metal foam}
\nomenclature[G, 03]{\(\mu\)}{Dynamic viscosity}
\nomenclature[G, 04]{${\alpha }_{\mathrm{sf}}$}{Specific surface area}
\nomenclature[G, 05]{\(\beta\)}{Thermal expansion coefficient}
\nomenclature[G, 06]{$\lambda$}{Thermal conduction coefficient}
\nomenclature[G, 07]{$\lambda_{\mathrm{b},x}$}{Thermal conductivity of battery in \textit{i}-direction}
\nomenclature[G, 08]{$\gamma$}{Liquid faction}

\nomenclature[X, 01]{b}{Battery}
\nomenclature[X, 02]{m}{Metal foam}
\nomenclature[X, 03]{PCM}{Phase change material}
\nomenclature[X, 04]{cha}{Charge}
\nomenclature[X, 05]{min}{Minimum}
\nomenclature[X, 06]{max}{Maximum}
\nomenclature[X, 07]{am}{Ambient}


\printnomenclature

\section*{Declaration of Competing Interest}
The authors declare that they have no known competing financial interests or personal relationships that could have appeared to influence the work reported in this article.


\section*{Appendix. Metal foam-embedded PCM between battery cells}
As mentioned in Section 4.1 of the manuscript, the integration of foam-embedded PCM between the cooling plate and each battery cell improves the cooling process during the discharge phase. However, when a metal foam-PCM layer is introduced between battery cells, due to the release of latent heat to the battery cells during standby, the temperature of the batteries will increase. In other words, solidification of the PCM injects a large portion of heat into the adjacent battery cells. Consequently, thermal runaway could occur in the battery pack. Therefore, the battery module configuration with the metal foam-PCM layer between the battery cells configuration was not discussed in the Results section. To gain insight into the cooling performance of battery modules with PCM foam components in the cooling plate and in between battery cells, this configuration is modelled for the discharge rate of \SI{3}{\crate}. The conceptual figure of this battery module configuration is shown in \textcolor{blue}{Fig.~\ref{fig:configs}a}. The ambient temperature and the coolant mass flow rate per channel are taken as \SI{25}{\celsius} and \SI{2.25}{\gram\per\second}, respectively. As shown in \textcolor{blue} {Fig.~\ref{fig:configs}b}, as 28 HC PCM foam is introduced between the batteries, the maximum temperature of the battery module decreases significantly. As an example, at $t = \SI{1200}{\second}$, $T_\mathrm{max}$ through a battery module with 28HC PCM foam in the cooling plate and between batteries is \SI{7.7}{\celsius} lower than the case with dual 28HC/35HC PCMs used only in the cooling plate (not between batteries). According to \textcolor{blue} {Fig.~\ref{fig:configs}c}, significant improvement in the temperature homogeneity through the battery module is resulted when 28HC PCM is used in between batteries and the cooling plate. As a result, installing foam-embedded PCM in between battery cells in conjunction with the cooling plate significantly improves the cooling process. However, since the PCM component is adjacent to the batteries, the release of a high amount of latent heat during standby might trigger thermal runaway. This hinders the practical use of foam-PCM layers between batteries. Therefore, in the present work, foam-embedded PCMs are used only in the cooling plate.

\begin{figap*}
    \centering
    \includegraphics[width=1\textwidth]{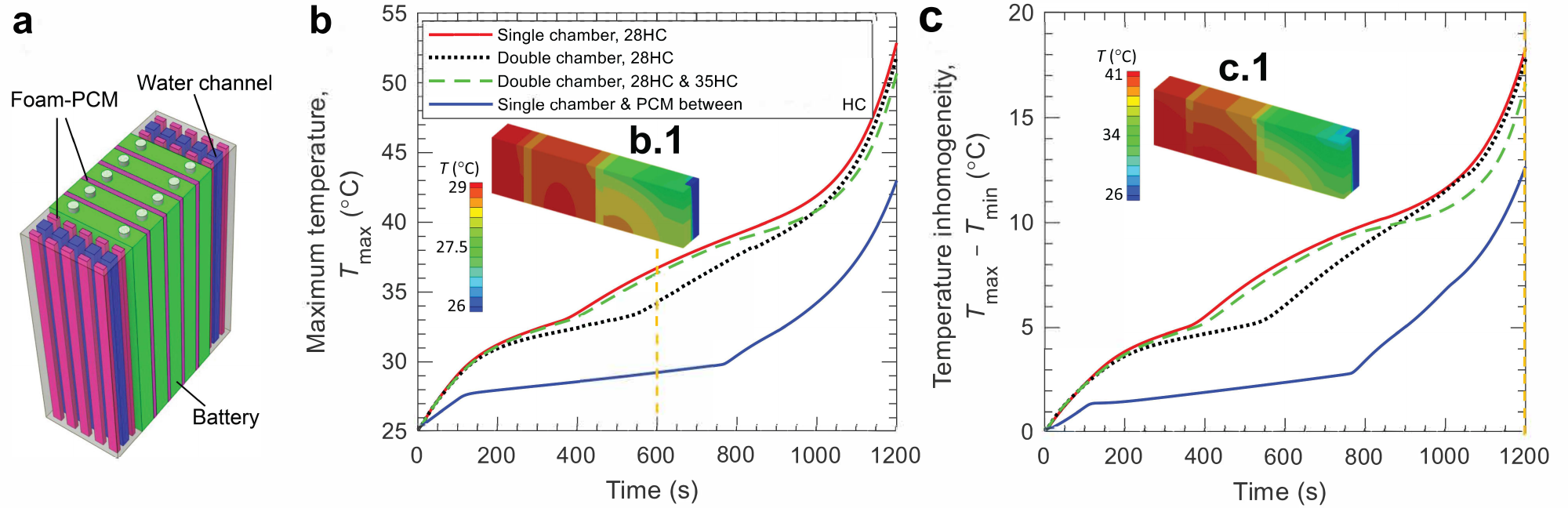}
    \caption{Cooling performance of four battery modules: single chamber (28 HC-foam), double chamber with 28HC PCM, double chamber with 28HC/35HC PCMs, and a combination of single chamber and foam-PCM in between batteries (28HC). The ambient temperature, coolant mass flow rate per channel and discharge rate are taken as \SI{25}{\celsius}, \SI{2.25}{\gram\per\second} and \SI{3}{\crate}, respectively.  
    \textbf{(a)}~Schematic of a battery pack with mono PCM-metal foam used in the cooling plate and in between batteries.
    \textbf{(b)}~Maximum temperature versus time through the modules. The inset illustrates the temperature contour at \SI{600}{\second} for the case where foam embedded PCM layers are installed between the batteries.
    \textbf{(c)}~Temperature inhomogeneity in the battery modules. The inset shows the temperature distribution at \SI{1200}{\second} for the module configuration in which foam-PCM is used between the batteries.}
\label{fig:configs}
\end{figap*}

\bibliographystyle{elsarticle-harv}
\bibliography{ms}


\end{document}